\documentclass{article}
\usepackage[numbers]{natbib}

\usepackage[final]{funasr_tech_report}

\usepackage{tabularx}
\usepackage[utf8]{inputenc} %
\usepackage[T1]{fontenc}    %
\usepackage{hyperref}       %
\usepackage{url}            %
\usepackage{booktabs}       %
\usepackage{amsfonts}       %
\usepackage{authblk} %
\usepackage{adjustbox}      %
\usepackage{nicefrac}       %
\usepackage{microtype}      %
\usepackage[table]{xcolor}  %
\usepackage{graphicx}
\usepackage{subcaption} %
\usepackage{amsfonts}       %
\usepackage{amsmath}        %
\usepackage{multicol}
\usepackage{multirow}
\usepackage{enumitem}
\usepackage{pifont}
\usepackage{array}
\usepackage{caption}

\usepackage[most]{tcolorbox}

\definecolor{sdPurple}{HTML}{7050D5}
\definecolor{sdTeal}{HTML}{168C91}
\definecolor{sdBlue}{HTML}{397ACA}

\newtcolorbox{sdexample}{
    enhanced,
    colback=white,
    colframe=sdPurple!35!white,
    boxrule=0.6pt,
    arc=3mm,
    left=3mm,
    right=3mm,
    top=2mm,
    bottom=2mm,
    title={Speaker-diarized transcription},
    fonttitle=\small\bfseries,
    coltitle=sdPurple!80!black,
    colbacktitle=sdPurple!7!white,
    titlerule=0pt
}

\newtcolorbox{sdsegment}[2]{
    enhanced,
    colback=#1!6!white,
    colframe=#1!25!white,
    boxrule=0.35pt,
    borderline west={2pt}{0pt}{#1},
    arc=1.5mm,
    left=3mm,
    right=3mm,
    top=2mm,
    bottom=2mm,
    before skip=2mm,
    after skip=2mm,
    fontupper=\small,
    before upper={%
        {\color{#1!80!black}\bfseries\ttfamily #2}\par
        \smallskip
        \raggedright
    }
}

\hypersetup{
    colorlinks=true,
    linkcolor=black,
    citecolor=black,
    urlcolor=QwenPurple
}

\usepackage{fontawesome5}

\definecolor{QwenPurple}{HTML}{7357D9}
\definecolor{QwenPurple}{HTML}{6F4FD8}
\definecolor{QwenPurpleLight}{HTML}{EEE9FF}
\definecolor{QwenLavender}{HTML}{F3F0FC}
\definecolor{ErrorRed}{HTML}{B54747}

\newcolumntype{L}[1]{%
  >{\raggedright\arraybackslash}p{#1}%
}

\newcommand{\bad}[1]{%
  \textcolor{ErrorRed}{\bfseries #1}%
}

\newcommand{\good}[1]{%
  \textcolor{QwenPurple}{\bfseries #1}%
}

\newcolumntype{Z}[1]{%
  >{\raggedright\arraybackslash}m{#1}%
}

\usepackage[utf8]{inputenc}
\usepackage[T1]{fontenc}
\usepackage{CJKutf8}

\title{Qwen-Audio-3.1-ASR Technical Report}

\author{
  \textbf{Alibaba Token Foundry, Alibaba Group}\\
  \href{https://qwenaudio.github.io/qwen-audio-3.1-asr/}{
    \faGlobe\enspace\textbf{Project Page}
  }
}

\begin{document}

\maketitle
\vspace{-3mm}
\begin{abstract}
We present \textbf{Qwen-Audio-3.1-ASR}, a family of audio-language models that builds upon its predecesor Qwen-Audio-3.0-ASR with improved speech recognition and expanded capabilities in speaker-diarized transcription and general audio understanding.
The family retains support for 30 languages and 16 Chinese dialects, together with instruction-controlled capabilities including historical-context conditioning, hierarchical hotword customization, and native transcription polishing.
Speaker-diarized ASR produces structured outputs containing speaker labels, timestamps, and recognized text, maintaining consistent speaker labels across turns and the representation of overlapping speech.
The general audio understanding capabilities cover audio-, speech-, and music-captioning, temporal grounding and sound event detection, as well as audio question answering and multiple-choice reasoning.
These tasks broaden the application scope from recognizing spoken words to interpreting speakers, acoustic events, and their temporal relations.
To accommodate different deployment requirements, the family includes variants for asynchronous file transcription, low-latency streaming, interactive voice input, and non-streaming recognition, with speaker-diarized transcription and general audio understanding available in certain variants.
Evaluations span conventional ASR, multilingual and dialectal speech, speaker-diarized transcription, and general audio understanding, demonstrating the capabilities of the Qwen-Audio-3.1-ASR family across transcription and broader audio analysis tasks.
\end{abstract}

\begin{figure*}[h]
    \centering
    \includegraphics[width=\textwidth]{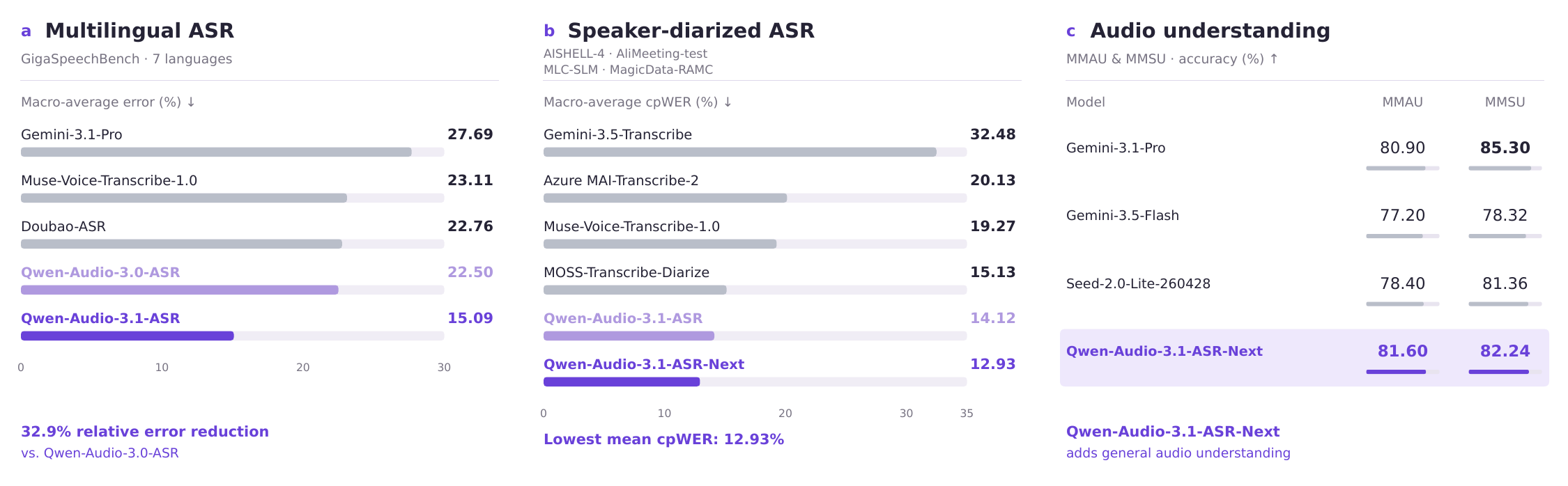}
    \caption{
    \textbf{Performance overview of Qwen-Audio-3.1-ASR on selected benchmarks.}
    (a) Multilingual ASR: unweighted mean of language-specific CER or WER (\%)
    across seven languages in GigaSpeechBench.
    (b) Speaker-diarized ASR: unweighted mean cpWER (\%) across AISHELL-4,
    AliMeeting-test, MLC-SLM, and MagicData-RAMC.
    (c) General audio understanding: accuracy (\%) on MMAU and MMSU,
    evaluated using the \textbf{Qwen-Audio-3.1-ASR-Next} variant.
    Lower is better in (a) and (b), and higher is better in (c). Full evaluation results are reported in Section~\ref{sec:exp}.
    }
    \label{fig:performance-overview}
\end{figure*}

\begin{figure}[t]
    \centering
    \includegraphics[width=\linewidth]{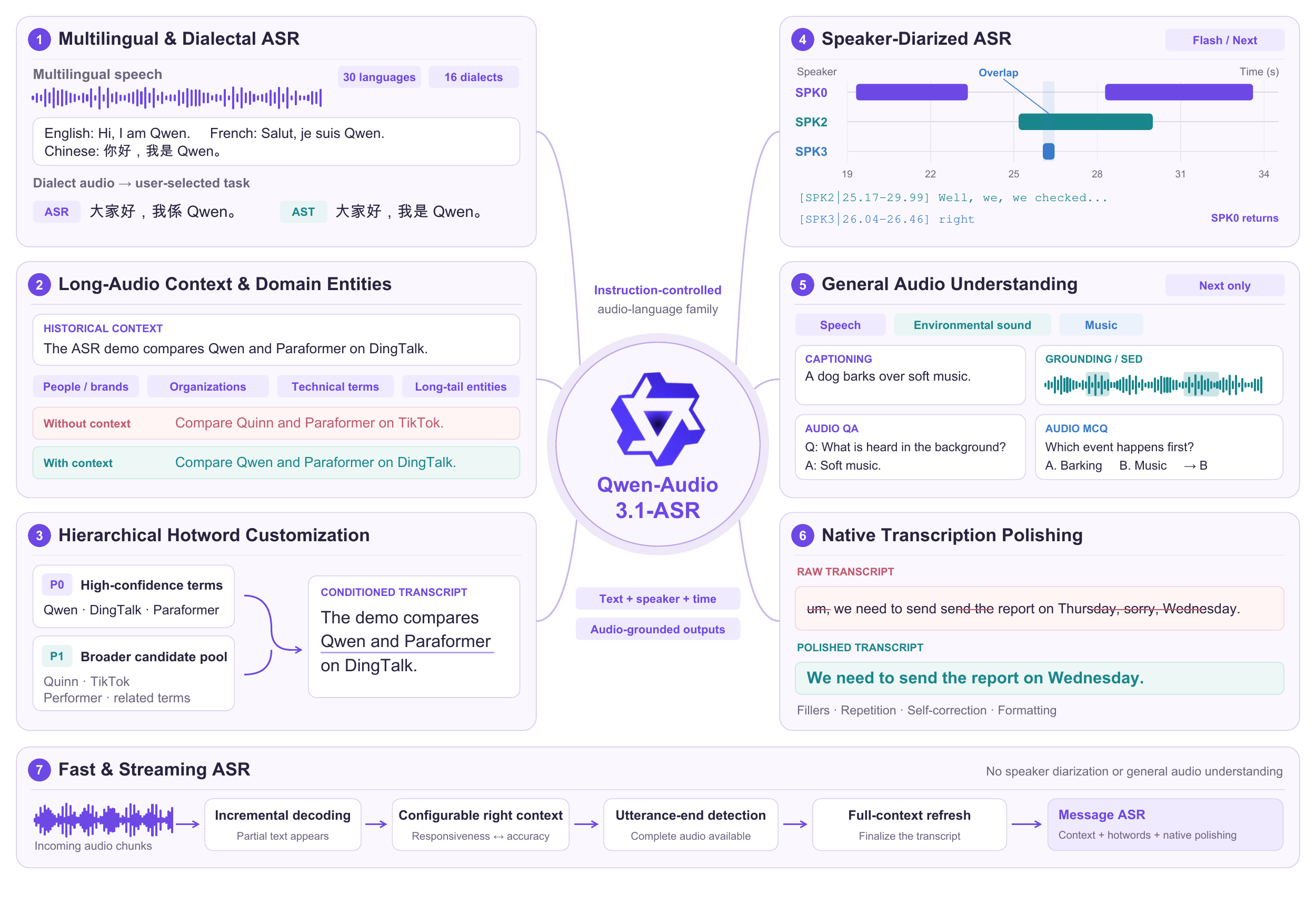}
    \caption{\textbf{Overview of Qwen-Audio-3.1-ASR and its capabilities.} The model family supports multilingual and dialectal speech recognition, user-selectable dialect transcription or translation into Chinese, long-audio context and domain entity recognition, hierarchical hotword customization, and native transcription polishing. Speaker-diarized ASR produces transcripts with speaker labels and timestamps, including overlapping speech, while general audio understanding supports captioning, temporal grounding, sound event detection, and audio question answering. Streaming recognition provides incremental transcripts for real-time display, followed by a non-streaming refresh at the utterance end. Capability availability depends on the model variant; streaming does not support speaker diarization or general audio understanding.}
    \label{fig:overview}
\end{figure}

\section{Introduction}
\label{sec:introduction}

Automatic speech recognition (ASR) provides the linguistic foundation for voice interfaces, meeting transcription, and conversational systems. However, a transcript alone only captures partial information in an audio recording. Multi-party conversations require identifying who spoke and when, while interpreting speech, environmental sounds, and music requires information that may not be recoverable from words alone. Practical audio systems must therefore address three complementary
questions: \emph{what was said}~\citep{DBLP:conf/icml/RadfordKXBMS23,q31_he2025cmtllm},
\emph{who said it and when}~\citep{q31_peng2026vibevoice},
and \emph{what the audio conveys beyond its transcript}~\citep{q31_sakshi2024mmau,q31_wang2025mmsu}. These requirements motivate extending accurate speech recognition toward structured transcription and general audio understanding.

Recent audio-language models have made substantial progress toward this broader scope. Qwen3-ASR demonstrates the potential of building multilingual speech recognition upon an audio-capable language model~\citep{q31_shi2026qwen3asr}. In parallel, Qwen3-Omni integrates audio perception into a broader multimodal framework~\citep{q31_xu2025qwen3omni}, while MOSS-Audio~\citep{q31_yang2026mossaudio} explicitly supports captioning, timestamped transcription, and time-aware audio understanding. More recently, FireRedAudio~\citep{q31_shen2026fireredaudio} combines recognition and audio understanding with speech generation in a general-purpose audio-language model. Together, these developments establish audio-conditioned language modeling as a useful framework for tasks that range from lexical recognition to descriptions and answers grounded in acoustic input.

Speaker-diarized transcription introduces a distinct structural requirement. Beyond recognizing individual utterances, a system must associate words with consistent speaker labels and temporal boundaries, including during rapid turn-taking and overlapping speech. Recent systems such as MOSS Transcribe Diarize~\citep{q31_mosi2026mosstd} and VibeVoice-ASR~\citep{q31_peng2026vibevoice} jointly model transcription, speaker diarization, and timestamp prediction, offering an alternative to assembling these outputs through separate processing stages. This direction is particularly relevant to meetings and interviews, where a lexically accurate transcript can still misrepresent a conversation if speaker assignments or temporal relations are incorrect.

General audio understanding further broadens the information that a model must preserve. Environmental acoustic events, vocal expressions, and musical characteristics can be central to audio comprehension even when they cannot be captured by a verbatim transcript. Moreover, identifying an event is not equivalent to locating it in time or reasoning about its relations to other sounds. Benchmarks such as MMAU~\citep{q31_sakshi2024mmau} evaluate understanding and reasoning across speech, sound, and music, while MMSU~\citep{q31_wang2025mmsu} examines fine-grained spoken-language understanding involving semantic, paralinguistic, and phonological information. These complementary perspectives motivate evaluating audio models beyond transcription error rates alone.

The challenge for a production-oriented model family is to develop these capabilities while retaining reliable recognition and practical control over its outputs. Multilingual and dialectal speech, specialized terminology, and evolving conversational context continue to put demands on ASR accuracy. Speaker-aware outputs additionally require consistent speaker labeling and accurate timing, whereas audio understanding requires responses grounded in the recording rather than plausible text alone. Deployment also imposes different constraints: low-latency streaming, asynchronous file transcription, and richer non-streaming analysis need not expose identical task capabilities. We therefore treat \textbf{recognition quality}, \textbf{output structure}, \textbf{audio understanding}, and \textbf{deployment configuration} as related but distinct aspects of the system.

In this report, we present \textbf{Qwen-Audio-3.1-ASR}, a family of audio-language models that builds upon Qwen-Audio-3.0-ASR~\citep{qwen_audio_3_0_asr} and introduces two major capability extensions: \textit{speaker-diarized ASR} and \textit{general audio understanding}. The family retains support for 30 languages and 16 Chinese dialects, together with production-oriented controls such as contextual conditioning, hierarchical hotword customization, and native transcription polishing in the corresponding deployment profiles. Qwen-Audio-3.1-ASR supports speaker-diarized transcription, producing recognized text with recording-local speaker labels and timestamps. Its Next variant \textbf{Qwen-Audio-3.1-ASR-Next} additionally supports general audio understanding. Speaker-diarized transcription and general audio understanding are available through the corresponding non-streaming interfaces, but not through the Streaming or Message ASR profiles. Figure~\ref{fig:performance-overview} summarizes representative results on multilingual ASR, speaker-diarized transcription, and general audio understanding, highlighting improved multilingual recognition over Qwen-Audio-3.0-ASR and the capabilities introduced in the Qwen-Audio-3.1-ASR family.

The main advances of this generation are summarized as follows:
\begin{itemize}[leftmargin=*,noitemsep]
    \item \textbf{Enhanced speech recognition.} We strengthen the core ASR capability across multilingual, dialectal, and application-oriented recognition settings, building on the instruction-controlled recognition framework of the previous generation. Contextual conditioning, hotword customization, and transcription polishing remain complementary controls rather than substitutes for accurate recognition.

    \item \textbf{Speaker-diarized ASR.} We extend transcription to structured outputs containing recognized text, speaker labels, and timestamps. The capability targets multi-speaker recordings with speaker changes, short conversational responses, and overlapping speech, preserving conversational structure beyond a flat sequence of words.

    \item \textbf{General audio understanding.} In addition to transcription, the Next variant is trained on captioning tasks spanning speech, environmental sounds, and music, as well as temporal grounding and audio question answering in both open-ended and multiple-choice formats, achieving comprehensive audio understanding.
\end{itemize}

Our evaluations span speech recognition, speaker-diarized transcription, and audio understanding, using task-appropriate evaluation protocols to evaluate these different capabilities. The remainder of this report describes the model architecture in Section~\ref{sec:architecture}, training in Section~\ref{sec:training}, deployment variants in Section~\ref{sec:deployment}, and experiments in Section~\ref{sec:exp}.

\section{Model Architecture}
\label{sec:architecture}

\begin{figure}[h]
    \centering
    \includegraphics[width=1.0\linewidth]{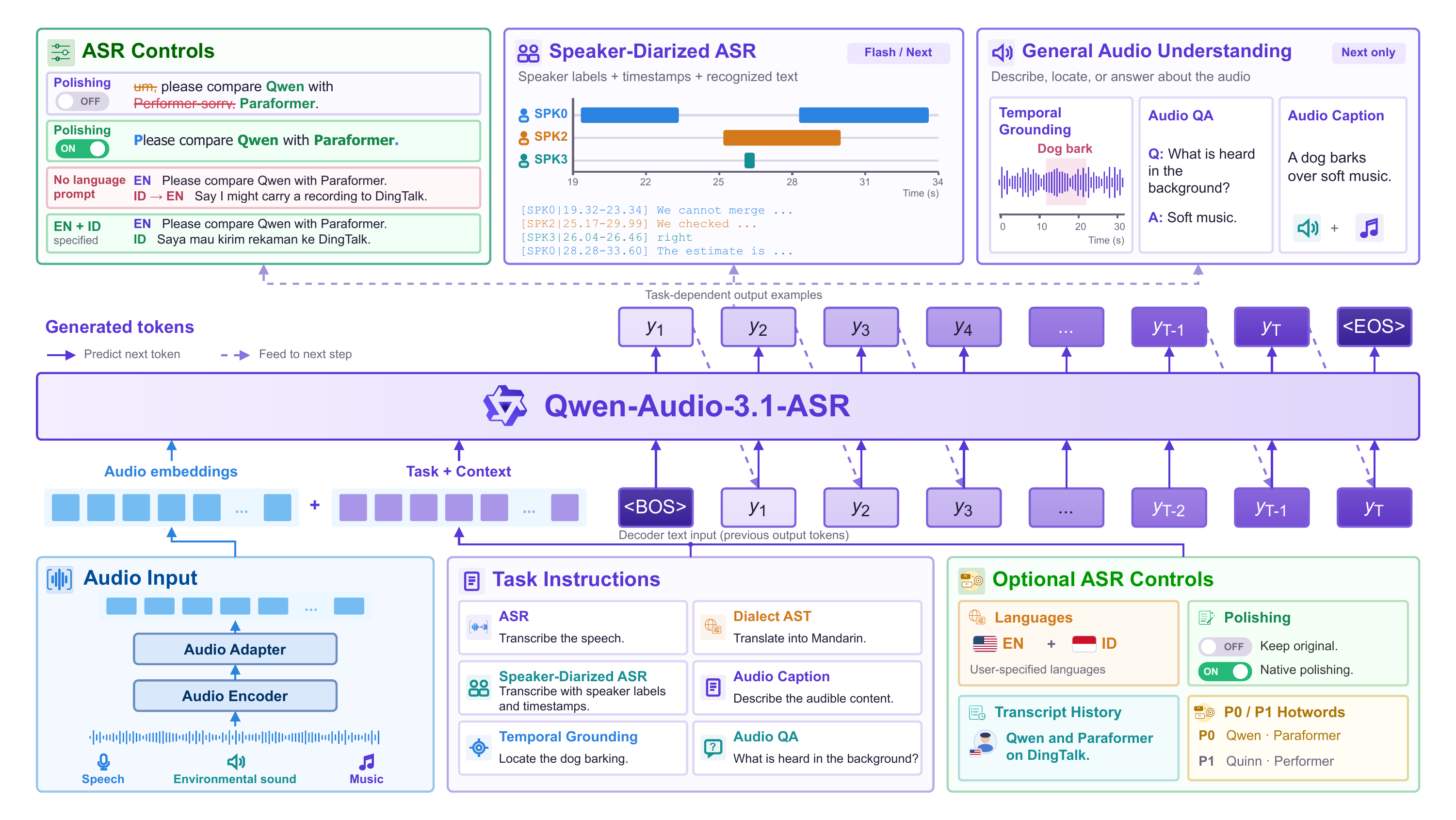} 
    \caption{Overview of the instruction-controlled decoding workflow of Qwen-Audio-3.1-ASR.}
    \label{fig:model}
\end{figure}

Qwen-Audio-3.1-ASR builds upon the audio-language modeling framework of Qwen-Audio-3.0-ASR and extends its output space from speech transcription to speaker-diarized transcription and general audio understanding. As illustrated in Figure~\ref{fig:model}, the core framework consists of an audio encoder, an audio adapter, and an LLM-based decoder. Audio representations provide the acoustic evidence, while task instructions and optional textual context specify the desired output behavior. An auxiliary CTC branch, omitted from the figure for clarity, provides preliminary recognition hypotheses to support contextual processing. The available model capabilities depend on the model variant and deployment profile.

\subsection{Audio Encoding and Instruction-Conditioned Decoding}
\label{subsec:audio-language-architecture}

The audio encoder transforms the input waveform into a sequence of acoustic representations. The audio adapter bridges these representations to the input space of the LLM decoder, allowing the decoder to generate outputs conditioned on both audio input and textual instructions. Unlike a transcript-only interface, this formulation retains direct access to the acoustic input when producing speaker-aware transcripts, audio descriptions, or answers to audio-related questions.

Let $x$ denote the input audio, $q$ the task instruction, and $c$ the optional textual conditions. The conditional generation process can be represented as
\begin{equation}
    H = A(E(x)), \qquad
    p_{\theta}(y \mid x,q,c)
    =
    \prod_{t=1}^{T}
    p_{\theta}(y_t \mid y_{<t},H,q,c),
\end{equation}
where $E$ is the audio encoder, $A$ is the audio adapter, and $H$ is the resulting sequence of audio embeddings. The model parameters are denoted by $\theta$. The task-dependent output is represented as $y=(y_1,\ldots,y_T)$, where $T$ is the output sequence length, $y_t$ is the token at position $t$, and $y_{<t}=(y_1,\ldots,y_{t-1})$ is the preceding output prefix.

For ASR, the textual conditions can include target languages, historical transcripts, hierarchical hotword candidates, and a polishing instruction. These controls specify how the recognized content should be rendered or contextualized. For audio understanding, the instruction instead specifies a description, a target event, or a question to be answered from the recording.

The auxiliary CTC branch operates on the audio encoder representations and provides an initial recognition hypothesis for hotword customization and a coarse alignment signal for long-audio context injection, as described in Sections~\ref{subsec:hotword} and~\ref{subsec:long-audio-context}. Its role is auxiliary: CTC hypotheses are not the sole representation of the recording, and should not be confused with the speaker-diarized transcriptions with timestamps or event boundaries produced for the corresponding tasks.

\subsection{Speaker-Diarized Transcription}
\label{subsec:speaker-diarized-output}

Speaker-diarized ASR extends the transcription target to include speaker labels and temporal information. Instead of returning a single undifferentiated text sequence, the model produces a structured transcript that associates each speech segment with its recognized content, speaker label, and temporal boundaries.

At the task level, a speaker-diarized transcript can be represented as
\begin{equation}
    \mathcal{S}
    =
    \left\{
        \left(b_i,e_i,k_i,w_i\right)
    \right\}_{i=1}^{M},
\end{equation}
where $\mathcal{S}$ denotes the structured transcript and $M$ is the number of speech segments. For segment $i$, $b_i$ and $e_i$ are the segment start and end times, $k_i$ is a speaker label, and $w_i$ is the corresponding transcription. This notation specifies the semantic content of the output rather than an exact implementation format. Speaker labels identify distinct speakers within a recording; they do not imply recognition of a person's real-world identity.

The output must preserve speaker consistency across turns while accommodating short responses and overlapping speech. In particular, temporal intervals associated with different speakers need not be disjoint. Jointly representing text, speaker-diarization labels and timing allows the transcript to retain conversational structure that would otherwise be lost in plain-text ASR output.

\subsection{General Audio Understanding}
\label{subsec:general-audio-output}

The Next variant broadens instruction-conditioned generation beyond lexical transcription to comprehensive audio understanding across speech, environmental sounds, and music. Specifically, its training incorporates three complementary task families within a unified framework: audio captioning, temporal grounding, and audio question answering.
\paragraph{Audio captioning.}
Audio captioning provides natural-language descriptions of audio content and its associated attributes. Depending on the audio type, these attributes may include speaker and paralinguistic characteristics, sound sources and acoustic environments, or musical properties such as genre, instrumentation, rhythm, and mood.

\paragraph{Temporal grounding.}
Temporal grounding associates a semantic query, such as a fine-grained caption or a sound event category, with its corresponding temporal locations in audio. We consider two grounding tasks: \textit{temporal boundary localization} predicts the start and the end boundaries of a single target interval, whereas \textit{temporal retrieval} identifies all time intervals that match the query. This formulation provides a unified view of single-interval localization and multi-interval retrieval.

\paragraph{Audio question answering.}
Audio question answering requires the model to interpret acoustic evidence in response to a given query. Both open-ended and multiple-choice formats are included, supporting the integration of audio perception with semantic reasoning.

\subsection{Deployment Variants}
\label{subsec:architecture-variants}

Table~\ref{tab:deployment-variants} summarizes the five deployment variants and their supported capabilities. Flash is reported as Qwen-Audio-3.1-ASR in the experimental results, while Next is reported as Qwen-Audio-3.1-ASR-Next. Deployment workflows are detailed in Section~\ref{sec:deployment}.

\begin{table*}[t]
    \centering
    \caption{Deployment variants of Qwen-Audio-3.1-ASR model family. $\checkmark$ indicates a supported capability; a dash indicates an unsupported capability. SD denotes speaker-diarized ASR with recording-local speaker labels and timestamps; AU denotes general audio understanding. Context, hotwords, and polishing refer to historical-context conditioning, hierarchical hotword customization, and native transcription polishing, respectively. SD and AU are available only in the corresponding non-streaming variants.}
    \label{tab:deployment-variants}
    \begingroup
    \small
    \setlength{\tabcolsep}{5pt}
    \renewcommand{\arraystretch}{1.35}
    \arrayrulecolor{black}
    \newcommand{\yes}{\textcolor{QwenPurple}{\checkmark}}
    \resizebox{\textwidth}{!}{
    \begin{tabular}{llcccccc}
        \toprule
        & & \multicolumn{6}{c}{Supported capabilities} \\
        \cmidrule(lr){3-8}
        \rowcolor{QwenPurpleLight}
        \textbf{Variant} & \textbf{Target scenario} &
        \textbf{ASR} & \textbf{SD} & \textbf{AU} &
        \textbf{Context} & \textbf{Hotwords} & \textbf{Polishing} \\
        \midrule
        Qwen-Audio-3.1-ASR-Flash-Filetrans & Asynchronous audio-file transcription &
        \yes & -- & -- & \yes & \yes & -- \\
        Qwen-Audio-3.1-ASR-Flash-Streaming & Real-time transcription &
        \yes & -- & -- & \yes & \yes & -- \\
        Qwen-Audio-3.1-ASR-Flash-Message & Interactive voice input &
        \yes & -- & -- & \yes & \yes & \yes \\
        \midrule
        Qwen-Audio-3.1-ASR-Flash & Non-streaming transcription &
        \yes & \yes & -- & \yes & \yes & \yes \\
        Qwen-Audio-3.1-ASR-Next & Non-streaming transcription and audio analysis &
        \yes & \yes & \yes & \yes & \yes & \yes \\
        \bottomrule
    \end{tabular}}
    \endgroup
\end{table*}

\section{Training}
\label{sec:training}

The training of Qwen-Audio-3.1-ASR comprises six stages: audio encoder pretraining, audio-language model pretraining, LLM cooldown adaptation, supervised fine-tuning, reinforcement learning, and multi-teacher on-policy distillation (MOPD). These stages progressively align acoustic representations with linguistic knowledge, develop instruction-following capabilities, and optimize task performance, followed by MOPD to consolidate capabilities of multiple teacher models into a single model.

\subsection{Pretraining of Audio Encoder}
\label{subsec:pretraining}
\begin{figure}[h]
    \centering
    \includegraphics[width=1\linewidth]{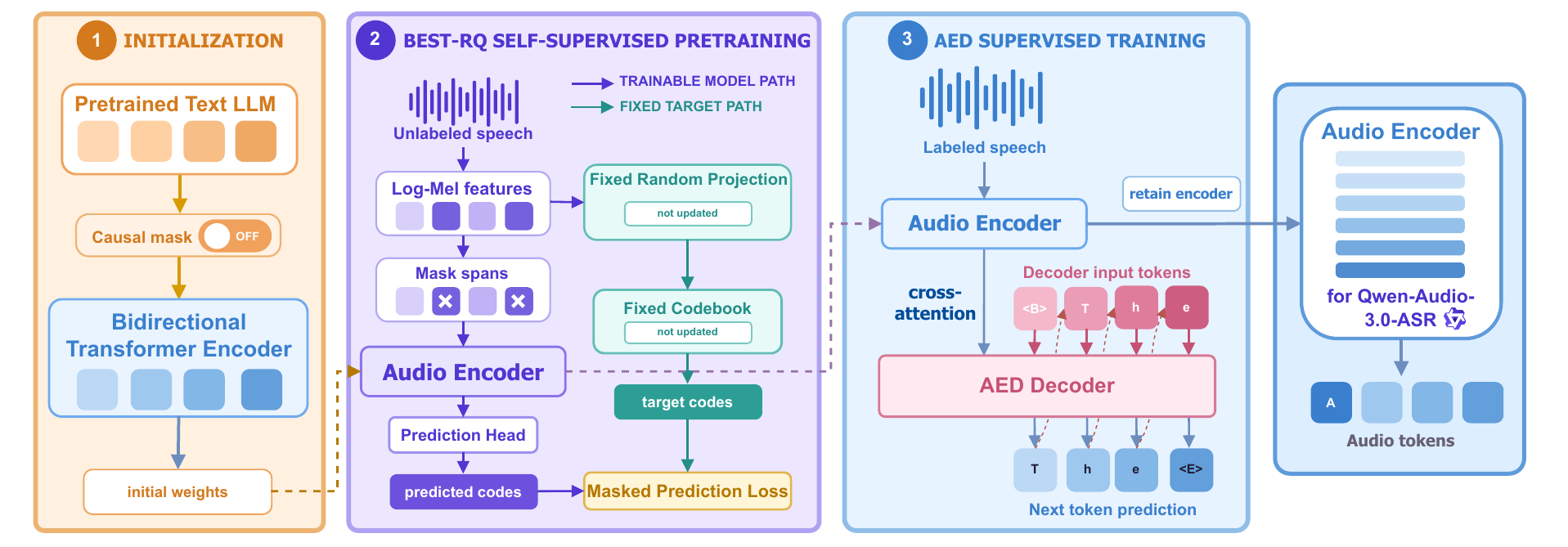}
    \caption{The pretraining pipeline for the audio encoder.}
    \label{fig:encoder}
\end{figure}

As illustrated in Figure~\ref{fig:encoder}, the audio encoder is trained in two consecutive stages: BEST-RQ self-supervised pretraining on large-scale unlabeled speech, followed by supervised training with an attention-based encoder--decoder (AED) objective on labeled audio--text data. Across the two stages, the pretraining corpus comprises tens of millions of hours of audio collected from diverse real-world scenarios.

\paragraph{BEST-RQ self-supervised pretraining.}
We first train the encoder using BEST-RQ~\citep{DBLP:conf/icml/ChiuQZYW22} on large-scale unlabeled speech collected from diverse real-world scenarios, covering domains such as artificial intelligence, biotechnology, e-commerce, education, entertainment, finance, and mobility. Acoustic frames are randomly masked, and the encoder predicts discrete targets generated from the corresponding unmasked features by a fixed random projection and a frozen codebook. This objective enables the encoder to learn contextual speech representations from large-scale and diverse unlabeled audio.
Following our previous study on cross-modal initialization
~\citep{DBLP:journals/corr/abs-2409-17750}, we initialize compatible Transformer layers of the encoder from a pretrained Qwen3 LLM~\citep{DBLP:journals/corr/abs-2505-09388}. For the non-streaming encoder, causal attention is replaced with bidirectional attention for speech encoding, while the acoustic front end is initialized separately. The resulting network is subsequently adapted to speech through BEST-RQ pretraining.

\paragraph{Supervised AED training.}
The pretrained encoder is then integrated into an AED model and optimized on large-scale labeled audio--text data, primarily covering Chinese and English. The labeled data are processed through a comprehensive pipeline that includes voice activity detection (VAD), pseudo-label generation by multiple ASR systems, including Paraformer-V2~\citep{DBLP:journals/corr/abs-2409-17746}, Whisper-large-v3\footnote{\url{https://huggingface.co/openai/whisper-large-v3}}, and SenseVoice~\citep{DBLP:journals/corr/abs-2407-04051}, followed by inverse text normalization (ITN).

Following the general training paradigm of SenseVoice-Large~\citep{DBLP:journals/corr/abs-2407-04051}, an autoregressive decoder attends to the encoder representations and predicts the reference transcription token by token, providing direct acoustic--text supervision. After this stage, the AED decoder is discarded, and the trained encoder is used to initialize the audio encoder of Qwen-Audio-3.1-ASR for subsequent Audio LLM pretraining.

\subsection{Pretraining of Audio LLM}
\label{subsec:pretraining-llm}
\paragraph{Initialization.}
The Audio LLM decoder is initialized from the language-model component of Qwen, while its vision-specific components are not used. The backbone adopts a Mixture-of-Experts (MoE) architecture. This initialization provides substantial linguistic capacity and multilingual knowledge while maintaining relatively low per-token computation and decoding cost, making it suitable for Qwen-Audio-3.1-ASR, including its latency-sensitive streaming variant.

\paragraph{Training data and objective.}
During Audio LLM pretraining, we train the model on approximately 20 million hours of multilingual data constructed through our large-scale data pipeline, consisting primarily of ASR data and interleaved text--audio data. The pretrained audio encoder is kept frozen, while the audio adapter and the LLM decoder are jointly optimized.
The model is trained with a cross-entropy loss over the target text tokens, enabling the audio adapter and LLM decoder to align acoustic representations with the LLM's textual representation space. This stage establishes the
fundamental audio--text alignment required for subsequent ASR adaptation.

The resulting Audio LLM combines the pretrained linguistic knowledge of Qwen with the audio--text alignment learned in this stage. It provides several useful priors for downstream ASR, including multilingual knowledge, world knowledge for resolving semantically ambiguous entities and homophones, and instruction-following capability for handling flexible textual conditions. 

\subsection{LLM Cooldown Adaptation}
\label{subsec:cooldown}

Following Audio LLM pretraining, we introduce a cooldown stage using approximately one million hours of higher-quality audio--text data. Compared with the preceding Audio LLM pretraining stage, this cooldown adaptation stage aims at refining audio--language alignment and emphasizes a more balanced distribution of languages and Chinese dialects, improved transcription quality, and a broader mixture of tasks and instructions before supervised fine-tuning. In addition to conventional ASR, the training mixture includes audio understanding, speaker diarization, and speech question-answering and dialogue data.

\paragraph{Data rebalancing and quality filtering.}
We rebalance the multilingual and dialectal ASR data to improve coverage across linguistic conditions. To further improve transcription quality, we re-transcribe the ASR data using VibeVoice and compare the resulting hypotheses with the existing reference transcripts. Samples with a word error rate greater than 0.3 are removed. This procedure uses agreement with an additional recognizer as a data-quality filter; it does not replace the reference transcripts with the re-recognition outputs.

\paragraph{Task diversification.}
The cooldown mixture broadens the training objectives beyond plain speech transcription. Audio understanding data introduce supervision for interpreting acoustic content, speaker diarization data expose the model to speaker-related structure, and speech question-answering and dialogue data enrich the range of audio-conditioned instructions and responses. Together with language and dialect rebalancing, this task mixture prepares the model for subsequent task-specific supervised fine-tuning.

The pretrained audio encoder remains frozen, while the audio adapter and LLM decoder are optimized using a cross-entropy objective over the target output tokens. 

\subsection{Supervised Fine-tuning (SFT)}
\label{subsec:SFT}

Supervised fine-tuning turns the integration of the pretrained audio encoder, audio adapter, and cooldown-adapted LLM into an instruction-controlled audio-language model. Building on the recognition capabilities of the previous generation Qwen-Audio-3.0-ASR, the SFT mixture additionally incorporates speaker-diarized ASR and general audio understanding data. The task mixture is configured according to the capabilities of the corresponding model variants.
During SFT, we jointly update the audio encoder and audio adapter, while keeping the base LLM parameters frozen and adapting the decoder through Low-Rank Adaptation (LoRA). Training examples follow a conversational format in which a user instruction and the associated audio are paired with an assistant response specifying the desired output. The model is optimized using an autoregressive cross-entropy objective over the target response.
The ASR portion retains broad coverage of multilingual, dialectal, code-switching, noisy, far-field, and application-oriented speech. Recognition instructions may specify target languages, historical transcript context, hierarchical hotword candidates, or transcription polishing. The additional speaker-diarized ASR and audio understanding examples extend this instruction-following interface to structured transcription and audio-conditioned semantic tasks.

\subsubsection{Recognition-oriented SFT}
\label{subsubsec:recognition-sft}

\paragraph{Specified-language transcription.}
To support flexible language control, the SFT mixture includes multilingual, dialectal, and code-switching speech covering the target languages and Chinese dialects. Each example may be accompanied by a list of target languages that specifies the expected transcription behavior. For monolingual transcription, the instruction requires the model to produce a transcript in the specified language, without translating the content into another language or introducing unintended cross-language substitutions. For multilingual speech, the model is instructed to preserve each supported language in its original written form. Training on diverse language combinations and code-switching examples enables the model to follow runtime language specifications and reduces unintended cross-language substitutions, particularly for lower-resource languages.

\paragraph{Domain-entity augmentation.} To improve recognition of domain-specific entities that are rare in general ASR data, we augment the SFT mixture with dedicated entity-focused examples. High-value entities are first mined from domains such as artificial
intelligence, finance, healthcare, and IT, and an LLM generates diverse short utterances containing these entities based on retrieved factual context. The generated texts are converted into speech using CosyVoice3~\citep{DBLP:journals/corr/abs-2505-17589}, supplemented by human recordings for terms that are difficult to synthesize reliably. The resulting audio is then re-decoded by the current ASR model, and samples with unambiguous entity-level mismatches are filtered out. These curated examples are incorporated into SFT to increase the model's exposure to rare and rapidly evolving domain terminology.

\paragraph{Native transcription polishing.}
A binary polishing instruction determines whether the target preserves the verbatim spoken form or produces a cleaned transcript. Polishing-enabled examples teach the model to remove non-semantic fillers, collapse stuttering and unnecessary repetitions, resolve explicit self-corrections, and normalize punctuation and textual formatting while preserving the speaker's final intent. The polishing data are constructed from both real and synthetic speech. For real speech, we retain the original audio from existing ASR records and regenerate the textual supervision using a text LLM with production polishing instructions, followed by fidelity filtering to remove samples with semantic loss or unsupported insertions. For synthetic data augmentation, an LLM injects fillers, repetitions, and self-corrections into canonical transcripts, and the resulting texts are converted into speech using TTS, followed by ASR-based pronunciation filtering and noise augmentation.
Since the polished transcript itself serves as the ASR target, recognition and polishing are learned within the same autoregressive decoding process, without requiring a cascaded ASR--LLM rewriting pipeline.

\paragraph{Historical-context conditioning.}
For contextual training, transcripts or interaction records from preceding turns are prepended as reference information. The instruction explicitly indicates that the historical context may be incomplete, irrelevant, or incorrect, and that the current audio should remain the primary source of evidence.
The context data are constructed from both real conversations and targeted synthetic examples. For real data, we match ASR records to their original sessions using audio identifiers, restore the chronological order of utterances, and use only preceding turns as context without LLM rewriting.
For synthetic augmentation, we construct examples involving homophone-ambiguous entities together with correct, conflicting, noisy, or empty historical contexts and hotword conditions.
This mixture encourages the model to exploit recurring entities and topic cues when they are informative, while avoiding unconditional copying from the provided context.

\paragraph{Hierarchical hotword conditioning.}
To train hierarchical hotword conditioning, each utterance is paired with two tiers of candidate hotwords. \textbf{P0} contains high-confidence, high-priority terms, including key named entities and other contextually critical expressions, whereas \textbf{P1} contains a broader candidate pool. An LLM assigns candidates to the two tiers by semantically matching them against the reference transcription, promoting unambiguous contextual matches
to P0 and retaining the remaining candidates in P1.
During SFT, both candidate sets are included in the instruction as auxiliary conditioning inputs. The model learns to place greater preference on P0 candidates when supported by the acoustic and contextual evidence, while
using P1 as a broader source of possible entities. This design improves long-tail hotword recall without forcing acoustically unsupported candidates into the transcript.

\subsubsection{Speaker-diarized ASR}
\label{subsubsec:speaker-diarized-sft}

\paragraph{Training data.}
The training data combine carefully annotated human transcripts with model-generated supervision. For film and television data, we clean the recordings and use their original subtitles together with multiple specialized models, including language identification, sound event detection, and speaker diarization models, to cross-validate the annotations and curate high-quality speaker-diarized ASR examples. We also construct simulated meeting conversations with LLM-generated dialogue topics and varied reverberation and noise conditions to broaden the coverage of conversational and acoustic scenarios.

\paragraph{Output serialization.}
Speaker-diarized ASR examples pair a multi-speaker recording with an instruction requesting both speaker diarization and transcription. The target response is serialized as newline-separated segments in the format \texttt{[SPK|start-end]text}. The example illustrates chronological serialization with overlapping time intervals and consistent speaker labels across turns.
\begin{sdexample}

\begin{sdsegment}{sdPurple}{[SPK0|19.36-23.36]}
We can't merge until we know who owns the label
definition for this release cycle.
\end{sdsegment}

\begin{sdsegment}{sdTeal}{[SPK2|25.20-30.00]}
Well, we, we checked the documentation, yeah,
and it is not clear at all.
\end{sdsegment}

\begin{sdsegment}{sdBlue}{[SPK3|26.08-26.48]}
right.
\end{sdsegment}

\begin{sdsegment}{sdPurple}{[SPK0|28.32-33.60]}
So the engineering estimate says it will take
three days to fix the schema.
\end{sdsegment}

\smallskip
{\footnotesize\color{black!60}
Segments are displayed in ascending order of the start time;
their time intervals may overlap.
SPK3's brief response overlaps with SPK2,
and SPK0 resumes before SPK2 finishes.
Repeated speaker labels retain the same color.
All timestamps are in seconds on an 80~ms grid.}

\end{sdexample}
Each segment contains a recording-local speaker label, a start--end time interval in seconds, and the corresponding recognized text. Repeated occurrences of the same speaker share the same label within the recording. The timestamps are expressed relative to the input audio, and intervals belonging to different speakers may overlap. The complete serialized response serves as the autoregressive training target, jointly supervising speaker labels, timestamps, and transcription text.

\begin{figure}[!htbp]
    \centering
    \includegraphics[width=\linewidth]{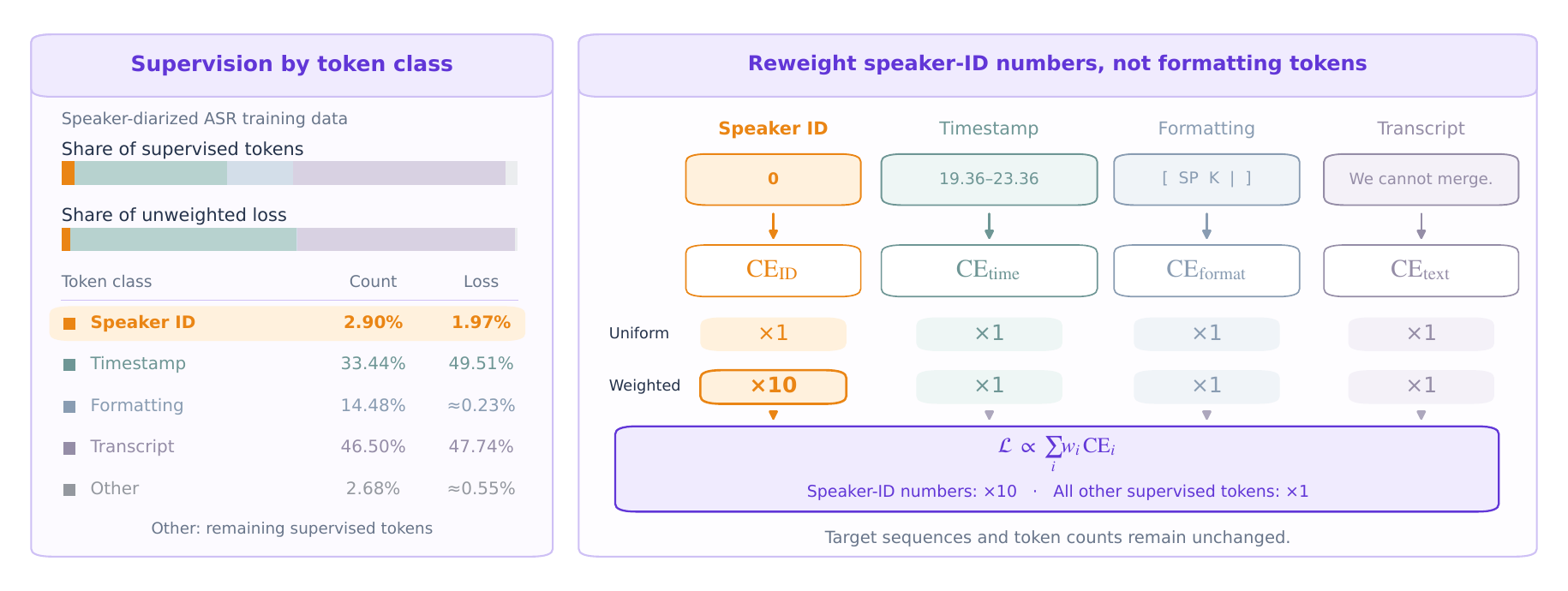}
    \caption{Speaker-token loss reweighting. Left: token-count and unweighted-loss shares for five categories in the speaker-diarized ASR training data. Other denotes supervised tokens outside the four named categories; its shares are calculated by subtraction. Formatting and other loss shares are approximate. Right: only speaker-ID number tokens receive a cross-entropy weight of 10; timestamp, formatting, transcription, and other supervised tokens retain a weight of 1. The example fields illustrate category membership rather than exact tokenizer segmentation. Target sequences and token counts remain unchanged. In the schematic loss expression, $i$ indexes supervised output tokens, $\mathrm{CE}_i$ is the cross-entropy loss for token $i$, and $w_i$ is its loss weight: 10 for speaker-ID number tokens and 1 for all other supervised tokens. The proportionality notation emphasizes relative token weights and omits loss normalization.}
    \label{fig:speaker-token-weighting}
\end{figure}

\paragraph{Speaker-token loss reweighting.}
To assess the balance of supervision across token categories, we examine token frequencies and loss contributions in the speaker-diarized ASR training data. Speaker-ID number tokens constitute 2.90\% of all supervised tokens but contribute only 1.97\% of the unweighted cross-entropy loss. Timestamp tokens account for 33.44\% of tokens and 49.51\% of loss, formatting tokens account for 14.48\% and approximately 0.23\%, and transcription tokens account for 46.50\% and 47.74\%, respectively. The remaining 2.68\% of tokens and approximately 0.55\% of loss are grouped as other supervised tokens, as shown in Figure~\ref{fig:speaker-token-weighting}.

These statistics reveal that speaker-ID prediction receives only a small share of the aggregate supervision despite its central role in speaker diarization. Timestamp prediction already accounts for a substantial fraction of the unweighted loss, while formatting is a relatively simple task that we observe the model already performs reliably. We therefore focus reweighting on speaker-ID prediction to better reflect its importance to diarization, rather than increasing the weight of every low-loss category. Specifically, we multiply the cross-entropy loss on speaker-ID number tokens by 10 while retaining unit weights for timestamp, formatting, transcription, and other supervised tokens. This adjusts the relative loss contributions without changing target sequences or token counts, and substantially improves both diarization error rate (DER) and concatenated minimum-permutation word error rate (cpWER) in our experiments.

\begin{figure}[!htbp]
    \centering
    \includegraphics[width=\linewidth]{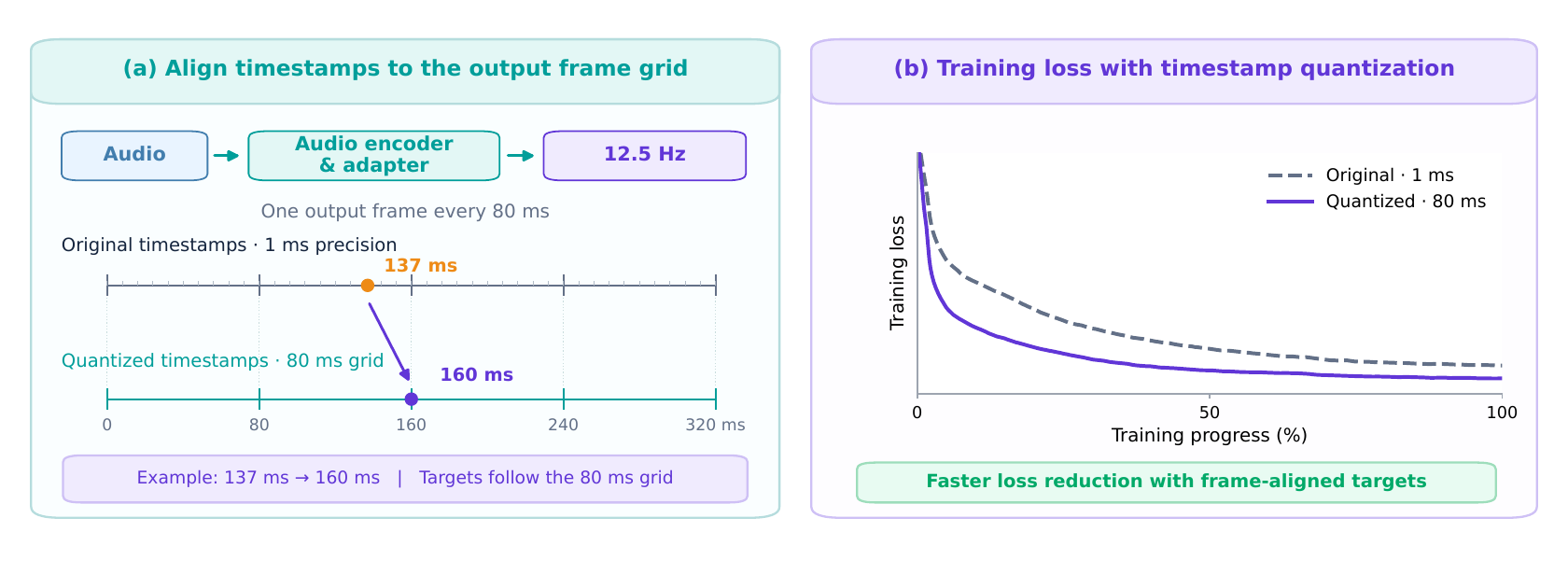}
    \caption{Frame-aligned timestamp quantization. (a) An illustrative timestamp is mapped from 1~ms precision to the 80~ms output frame grid of the audio encoder and adapter. (b) Training loss with original and quantized timestamps. Curves are digitized from training plots; absolute loss values and iteration counts are omitted. Both curves retain a shared vertical scale and use the same normalization of training progress.}
    \label{fig:timestamp-quantization}
\end{figure}

\paragraph{Frame-aligned timestamp quantization.}
We also find that the temporal resolution of the supervision materially affects optimization. Our initial targets retained the original timestamps at 1~ms precision. We instead quantize timestamps to an 80~ms grid, matching the 12.5~Hz frame rate of the audio representations produced by the audio encoder and adapter (Figure~\ref{fig:timestamp-quantization}). Compared with retaining the original 1~ms precision, this quantization yields substantially faster convergence of the training loss and of both DER and cpWER, as well as better DER and cpWER under the same training-data mixture. Together, speaker-token loss reweighting and frame-aligned timestamp quantization improve learning of the speaker and temporal structure represented in the serialized output.

\subsubsection{General Audio Understanding}
\label{subsubsec:audio-understanding-sft}
For the Next variant, the SFT mixture is augmented with audio-understanding examples spanning speech, environmental sounds, and music. These examples follow the same conversational format as the transcription data, with each audio input paired with a task-specific instruction and a reference response.

The additional data cover three task families. Audio captioning examples supervise natural-language descriptions of audio content and its acoustic attributes. Temporal grounding examples pair fine-grained audio descriptions or sound-event queries with their annotated temporal intervals. Audio question-answering examples pair questions about the recording with reference answers in both open-ended and multiple-choice formats.

For all tasks, the model is trained to autoregressively predict the reference response conditioned on the audio and task instruction. This unified formulation extends supervision beyond lexical transcription to semantic descriptions, temporal annotations, and audio-grounded answers, while allowing the instruction to specify the required output form.

\subsection{Reinforcement Learning (RL)}
\label{subsec:RL}

Following SFT, we apply reinforcement learning to optimize Qwen-Audio-3.1-ASR for transcription accuracy and speaker-diarized ASR. RL complements token-level maximum-likelihood training by directly optimizing sequence-level measures of recognition quality, speaker diarization, and temporal consistency.

\paragraph{RL data.}
We construct a compact, high-quality RL dataset focusing on challenging real-world cases. The ASR portion includes: (1) hard cases identified from disagreements between the base model and multiple external ASR systems, including Whisper, FireRed-ASR, and SenseVoice; (2) long-duration utterances exceeding 20 seconds; (3) hallucination-related samples, including both actual hallucinations and genuine long repetitions that may superficially resemble hallucinations; (4) keyword- and hotword-focused samples; and (5) regular ASR data included to mitigate catastrophic forgetting. Context-conditioned examples retain their associated historical context and instructions so that reward computation reflects the requested transcription behavior.

For speaker-diarized ASR, training examples pair multi-speaker recordings with reference transcripts, speaker labels, and segment timestamps. The structured references support joint evaluation of what is said, who says it, and when it occurs, using the output serialization described in Section~\ref{subsubsec:speaker-diarized-sft}. Each example retains its task identity and conditioning information, allowing the corresponding reward function to be selected for each generated response.

\paragraph{The FunVerl framework.}
We develop \textbf{FunVerl}, an audio-aware, fully asynchronous reinforcement learning framework built on verl~\citep{sheng2024hybridflow}. FunVerl extends the training infrastructure with audio preprocessing, multimodal rollout generation, and task-specific reward computation. Its asynchronous execution decouples response generation from policy optimization, allowing these workloads to run concurrently on separate resources.

\begin{figure}[htbp]
    \centering
    \includegraphics[width=1.0\linewidth]{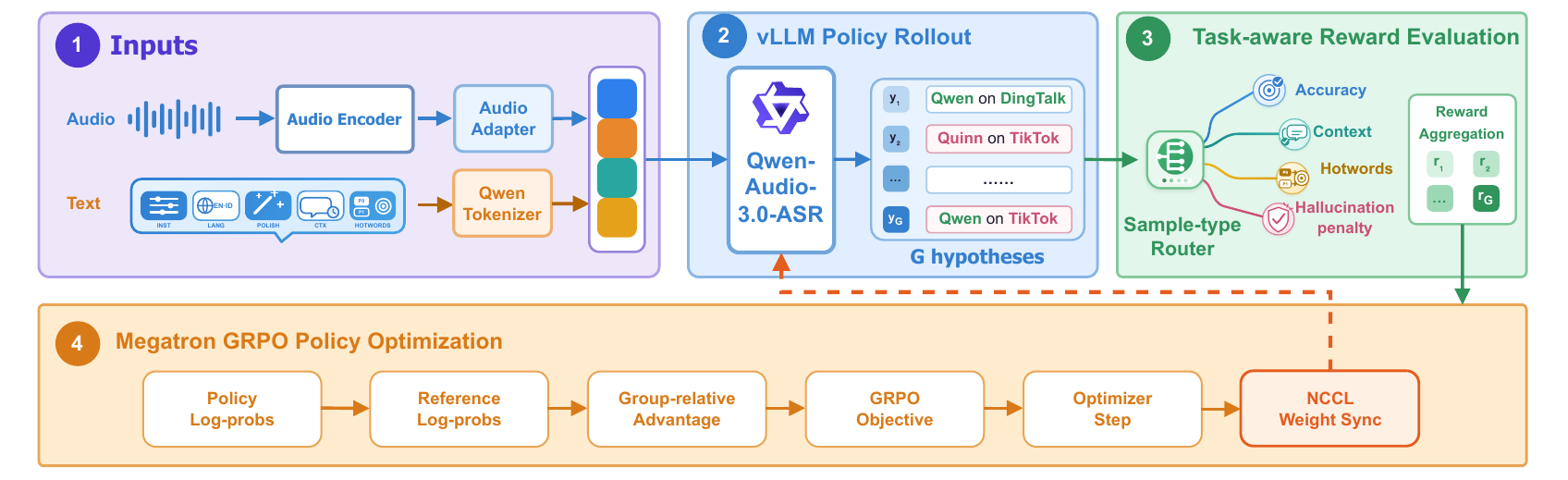}
    \caption{The FunVerl asynchronous RL framework. Audio preprocessing and vLLM rollout generation produce candidate responses, task-specific rewards score the responses, and the Megatron trainer updates the policy using GRPO. A trajectory queue decouples generation from optimization, and updated policy weights are synchronized to the rollout engine.}
    \label{fig:funverl}
\end{figure}

As illustrated in Figure~\ref{fig:funverl}, FunVerl consists of four modules orchestrated by Ray:

\begin{itemize}[leftmargin=*,noitemsep]
\item \textbf{Audio processor.} Raw waveforms are converted into Mel-spectrogram features, while task instructions, context, and hotwords are tokenized. Audio features and text tokens are passed together to the rollout engine.

\item \textbf{vLLM rollout engine.} The audio encoder and adapter transform Mel features into frame-level embeddings, which are combined with text embeddings. The MoE decoder then samples a group of $G$ candidate responses for each audio--instruction pair.

\item \textbf{Reward module.} Generated responses are scored using the reward associated with each sample's task and data source. ASR rewards evaluate transcription and instruction-conditioned behavior, while speaker-diarized ASR rewards jointly assess speaker diarization, transcription accuracy, and timing.

\item \textbf{Megatron policy trainer.} Megatron-Core supports policy optimization with tensor, pipeline, and expert parallelism. Updated policy weights are synchronized to the rollout workers through an NCCL-based weight-transfer mechanism, avoiding full model reloading between updates.
\end{itemize}

The rollout engine pushes generated trajectories into a queue, while the trainer independently consumes scored batches and updates the policy. This overlaps generation, reward computation, and optimization. A policy-version staleness threshold excludes excessively outdated trajectories and limits the lag between the rollout and training policies.

\paragraph{Policy optimization.}
We employ Group Relative Policy Optimization (GRPO)~\citep{shao2024deepseekmath} for both tasks. Given an audio--instruction input $q$, a rollout policy $\pi_{\theta_{\mathrm{old}}}$ generates responses $\{o_i\}_{i=1}^{G}$ with task-specific rewards $\{R_i\}_{i=1}^{G}$. The rewards are normalized within each group to obtain an advantage shared by the response tokens, without requiring a separately trained value network:

\begin{equation}
\hat{A}_{i,t}=\frac{R_i-\operatorname{mean}\!\left(\{R_j\}_{j=1}^G\right)}{\operatorname{std}\!\left(\{R_j\}_{j=1}^G\right)+\delta},
\end{equation}
where $\delta>0$ is a small constant for numerical stability. The policy maximizes the following group-level clipped surrogate objective with a KL regularizer:

\begin{equation}
\begin{aligned}
J_{\mathrm{GRPO}}(\theta)
&=\frac{1}{G}\sum_{i=1}^{G}\frac{1}{|o_i|}\sum_{t=1}^{|o_i|}
\Bigl[\min\!\bigl(r_{i,t}(\theta)\hat{A}_{i,t},\,
\operatorname{clip}(r_{i,t}(\theta),1-\varepsilon,1+\varepsilon)\hat{A}_{i,t}\bigr)\\
&\hspace{42mm}-\beta D_{\mathrm{KL}}\!\bigl(\pi_\theta(\cdot\mid h_{i,t})\|\pi_{\mathrm{ref}}(\cdot\mid h_{i,t})\bigr)\Bigr],
\end{aligned}
\end{equation}
where $h_{i,t}=(q,o_{i,<t})$, $r_{i,t}(\theta)=\pi_{\theta}(o_{i,t}\mid h_{i,t})/\pi_{\theta_{\mathrm{old}}}(o_{i,t}\mid h_{i,t})$, $\pi_{\mathrm{ref}}$ is a fixed reference policy, $\varepsilon$ is the clipping threshold, and $\beta$ controls KL regularization.

\subsubsection{ASR Reward}
\label{subsubsec:asr-reward}

For ASR, the reward combines transcription accuracy with criteria for practical recognition scenarios. Optimizing a WER-derived reward shares the task-level motivation of minimum word error rate (MWER) training. We retain the following reward components, selected according to the sample type:

\begin{itemize}[leftmargin=*,noitemsep]
    \item \textbf{ASR Accuracy ($R_i^1$).} $1 - \operatorname{WER}(y^*, y)$ as the primary reward, where $y^*$ and $y$ are the reference and hypothesized transcripts, respectively, and WER is expressed as a ratio.
    
    \item \textbf{Keyword Accuracy and Recall ($R_i^2$).} Keyword recall encourages correct entity recognition; keyword accuracy penalizes insertion errors to balance precision and recall. Keywords are either manually annotated or LLM-identified.
    
    \item \textbf{Hotword Reward ($R_i^3$).} For hotword-conditioned samples, a dedicated reward scores recognition of P0/P1 hotwords provided in the prompt, with asymmetric positive/negative gamma weighting.
    
    \item \textbf{Context Reward ($R_i^4$).} For context-conditioned samples, rewards hit/miss of context-derived entities with configurable penalties for false insertions.
    
    \item \textbf{Hallucination Suppression ($R_i^5$).} Hallucinated content is detected via pattern matching and penalized proportionally to the hallucinated span length.
    
\end{itemize}

FunVerl's reward routing mechanism selects the appropriate reward configuration per sample based on its data source (general, dialect, hotword, or context), enabling a single training run to jointly optimize across these recognition scenarios.

\subsubsection{Speaker-diarized ASR Reward}
\label{subsubsec:sd-reward}

For speaker-diarized ASR, we use three complementary reward components to evaluate transcription quality, speaker diarization, and temporal consistency:
\begin{itemize}[leftmargin=*,noitemsep]
    \item \textbf{Diarization Reward (DER).} A reward based on diarization error rate penalizes missed speech, false-alarm speech, and speaker confusion, encouraging accurate speaker activity and consistent speaker labeling across turns.
    \item \textbf{Transcription Reward (cpWER).} A reward based on concatenated minimum-permutation word error rate evaluates the transcript concatenated for each speaker under the best speaker permutation. It encourages accurate recognition and speaker assignment without depending on arbitrary speaker-ID names.
    \item \textbf{Temporal Consistency Reward (tcpWER).} A reward based on time-constrained cpWER additionally requires temporally compatible word alignments, encouraging recognized words to be associated with the correct time intervals.
\end{itemize}
Lower error rates correspond to better reward values. Together, these components encourage the model to produce accurate words, consistent speaker labels, and reliable timestamps within the same structured response.

\begin{figure}[htbp]
    \centering
    \includegraphics[width=\linewidth]{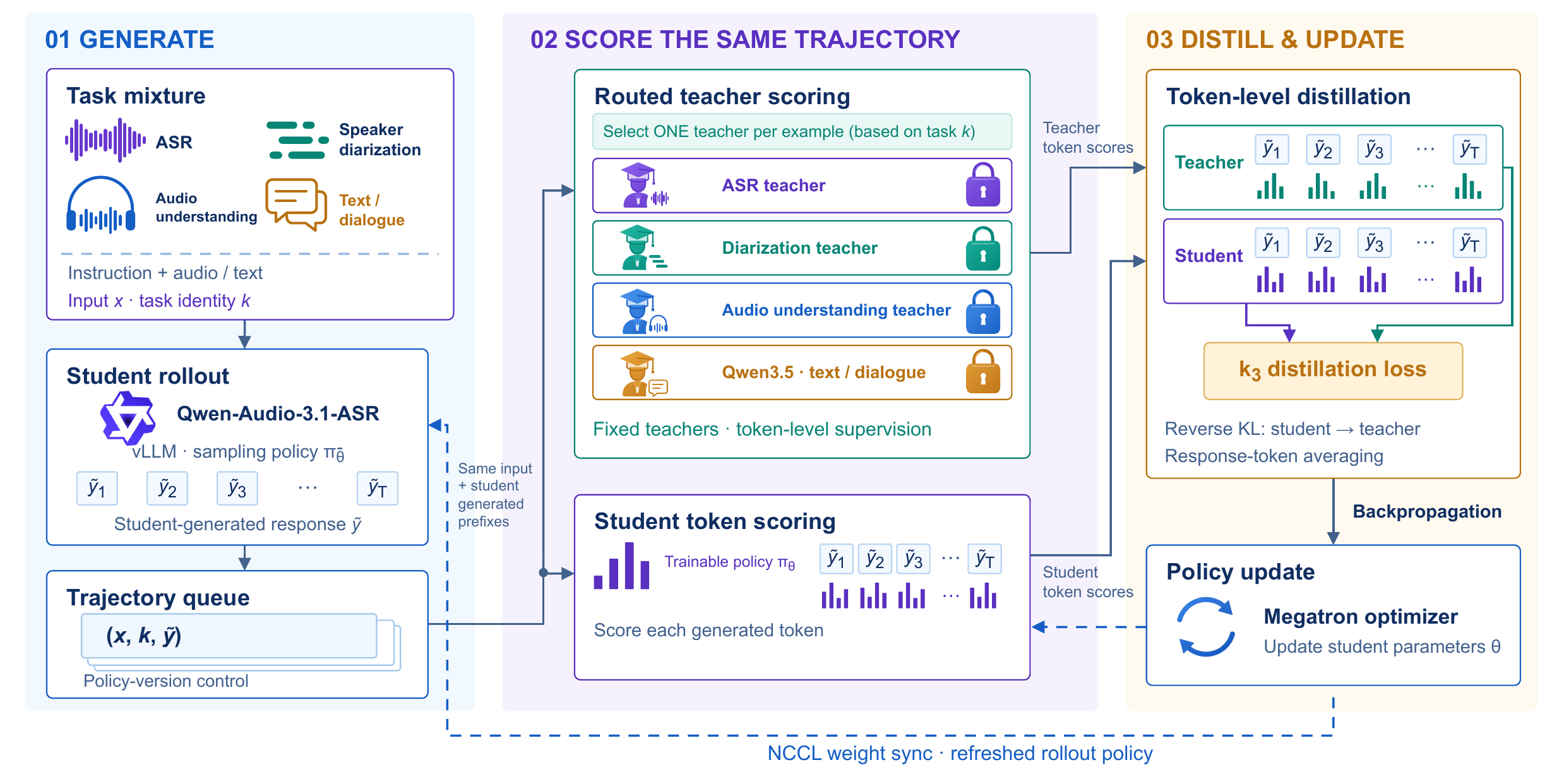}
    \caption{Overview of multi-teacher on-policy distillation in FunVerl.
    Student-generated responses are scored by task-routed teachers and
    the student using the same input and response prefixes.
    The $k_3$ objective provides token-level distillation supervision,
    while NCCL weight synchronization refreshes the rollout policy
    after student updates.}
    \label{fig:opd-framework}
\end{figure}

\subsection{On-policy Distillation}
\label{subsec:OPD}

As illustrated in Figure~\ref{fig:opd-framework}, we introduce
on-policy distillation (OPD) to consolidate complementary capabilities
from task-specialized teachers into a single audio-language student. Whereas RL supplies sequence-level feedback through task metrics, OPD supplies token-level feedback from a teacher on responses generated by the student. This provides a common training mechanism for transferring transcription, speaker-diarization, and audio-understanding capabilities.

\subsubsection{Learning from Student-generated Responses}
\label{subsubsec:opd-objective}

In conventional supervised distillation, the student is often trained on fixed reference responses or teacher-generated sequences. Such sequences can differ from the student's own predictions at inference time. OPD addresses this distribution mismatch by sampling responses from the current student and evaluating the teacher on the resulting student-generated prefixes. The teacher therefore provides guidance at the states that the student actually visits, including prefixes containing the student's own mistakes.

Let $x$ denote the task input, including the instruction and any audio or conversational context, and let $k$ identify the corresponding teacher. At each iteration, a student rollout policy $\pi_{\bar\theta}$ samples a response $\tilde y$. The fixed teacher $p_{\phi_k}$ then scores the sampled tokens conditioned on the same input and student-generated prefix $\tilde y_{<t}$.

We use the \textbf{$k_3$ estimator} of the reverse KL divergence from the student to the teacher. For each generated token, we define the teacher-to-student probability ratio and its distillation penalty as
\begin{equation}
\label{eq:opd-k3}
\rho_t(\theta)=\frac{p_{\phi_k}(\tilde y_t\mid x,\tilde y_{<t})}
{\pi_\theta(\tilde y_t\mid x,\tilde y_{<t})},
\qquad
d_t^{(k_3)}(\theta)=\rho_t(\theta)-\log\rho_t(\theta)-1.
\end{equation}
The $k_3$ penalty is nonnegative and uses teacher and student log-probabilities for the sampled token. Under on-policy sampling, its expectation estimates $D_{\mathrm{KL}}(\pi_\theta\|p_{\phi_k})$ at the corresponding prefix. The distillation objective averages these token-level penalties over student-generated responses:
\begin{equation}
\label{eq:opd-objective}
\mathcal{L}_{\mathrm{OPD}}(\theta)
=\mathbb{E}_{\substack{(x,k)\sim\mathcal{D}_{\mathrm{distill}}\\
\tilde y\sim\pi_{\bar\theta}(\cdot\mid x)}}
\left[\frac{1}{|\tilde y|}\sum_{t=1}^{|\tilde y|}d_t^{(k_3)}(\theta)\right],
\end{equation}
where $\mathcal{D}_{\mathrm{distill}}$ is the distribution of inputs paired with their assigned teachers. Teacher parameters and sampled trajectories are held fixed during each student update, and new trajectories are generated as the student evolves. FunVerl's rollout, training, and weight-synchronization components provide the underlying execution pipeline; in asynchronous execution, policy-version control limits the lag of student-generated trajectories.

\subsubsection{Teacher Models and Task Routing}
\label{subsubsec:opd-teachers}

We use three specialized audio teachers together with a Qwen3.5 teacher for text and dialogue. Examples are routed to the corresponding teacher according to the requested task:
\begin{itemize}[leftmargin=*,noitemsep]
    \item \textbf{ASR teacher.} The ASR teacher guides speech-to-text prediction, transferring its recognition capabilities to the student under the corresponding transcription instructions.
    \item \textbf{Speaker-diarization teacher.} The speaker-diarization teacher provides supervision for speaker-diarized transcription, including speaker labels, temporal boundaries, and the associated words. Its role is to guide the joint structured output, complementing the metric-based reward in Section~\ref{subsubsec:sd-reward}.
    \item \textbf{Audio understanding teacher.} The audio understanding teacher guides responses that require interpreting speech, environmental sounds, or music beyond literal transcription, such as audio descriptions and answers to audio-conditioned questions.
    \item \textbf{Text and dialogue teacher (Qwen3.5).} Qwen3.5 provides token-level supervision for text and dialogue examples, helping preserve the student's language modeling, instruction-following, and conversational capabilities during audio specialization.
\end{itemize}
Task-based routing lets each teacher supervise its corresponding data while contributing to the same student. The instruction determines whether the output is a transcript, a speaker-diarized transcript with timestamps, an audio-grounded answer, or a text response.

\subsubsection{Training Data Mixture}
\label{subsubsec:opd-data}

The training mixture combines \textbf{ASR}, \textbf{speaker diarization (SD)}, and \textbf{audio understanding (AU)} data with \textbf{text and dialogue data}. ASR examples cover speech transcription; SD examples retain speaker and timing structure; and AU examples retain the task instructions needed to interpret and reason about audio. The three audio data groups are paired with their corresponding specialized teachers, preserving their distinct output requirements within a shared training pipeline.

Text and dialogue examples are distilled from \textbf{Qwen3.5} using the same student-generated rollout and $k_3$ objective. For text examples, the input contains the textual prompt or instruction; dialogue examples additionally preserve the preceding conversational context. These data help retain general language and conversational capabilities while the audio data develop recognition, speaker diarization, and audio understanding. The resulting mixture supports both audio-grounded tasks and context-appropriate text responses within the same student.

\section{Deployment}
\label{sec:deployment}
Qwen-Audio-3.1-ASR provides deployment variants for file transcription, low-latency streaming, context-aware voice input, speaker-diarized transcription, and general audio understanding. The Streaming and Message ASR profiles retain the workflows of Qwen-Audio-3.0-ASR. The family additionally introduces explicit selection between dialect transcription and dialect-to-Mandarin translation, as well as speaker-diarized transcription and general audio understanding in the corresponding variants. These capabilities are variant-dependent rather than uniformly available across all deployment profiles.

\subsection{Multilingual and Dialectal ASR}
\label{subsec:multilinguality}

Qwen-Audio-3.1-ASR retains the instruction-controlled multilingual recognition interface of Qwen-Audio-3.0-ASR, supporting 30 languages and 16 Chinese dialectal varieties spanning eight major dialect regions. Applications can specify target languages at inference time to accommodate multilingual and code-switching speech without switching between language-specific checkpoints. For multilingual transcription, supported languages are preserved in their original writing systems.

A new capability in this generation is user-selectable output behavior for Chinese dialectal speech. Through an explicit task instruction, users can choose between ASR and automatic speech translation (AST):

\begin{itemize}[leftmargin=*,noitemsep]
    \item \textbf{Dialect ASR.} The model transcribes the dialectal speech while preserving its dialect-specific wording and expressions, rather than translating them into Mandarin.
    \item \textbf{Dialect-to-Mandarin AST.} The model translates the dialectal speech into written Mandarin, conveying the meaning of the utterance rather than preserving its original dialectal expressions.
\end{itemize}

This explicit choice separates faithful dialect transcription from dialect-to-Mandarin translation. It allows applications to select the appropriate output for scenarios such as dialect documentation or cross-dialect communication. The ASR/AST selection specifies the linguistic task and is distinct from transcription polishing, which controls the removal of disfluencies and the presentation of recognized text.

\subsection{Low-Latency Streaming Recognition}
\label{subsec:streaming recognition}

The streaming deployment profile retains the hybrid incremental and full-context recognition workflow of Qwen-Audio-3.0-ASR. Incoming audio is processed incrementally to provide provisional transcripts before an utterance is complete. After endpoint detection, the complete utterance is used to refresh and finalize the transcript. This workflow combines immediate textual feedback with recognition informed by the full utterance context.

Low-latency recognition is achieved by processing audio in short chunks with limited right context. Chunk size and right-context length provide configurable operating points for the latency--accuracy trade-off: shorter contexts prioritize responsiveness, while longer contexts provide additional acoustic evidence before the transcript is updated. Partial hypotheses may be revised as more audio becomes available.

The streaming variant supports incremental speech recognition but does not provide speaker-diarized transcription or general audio understanding. These capabilities are available through the corresponding non-streaming variants.

\subsection{Message ASR for Context-Aware Transcription}
\label{subsec:message asr}

\begin{figure*}[t]
    \centering
    \includegraphics[width=\textwidth]{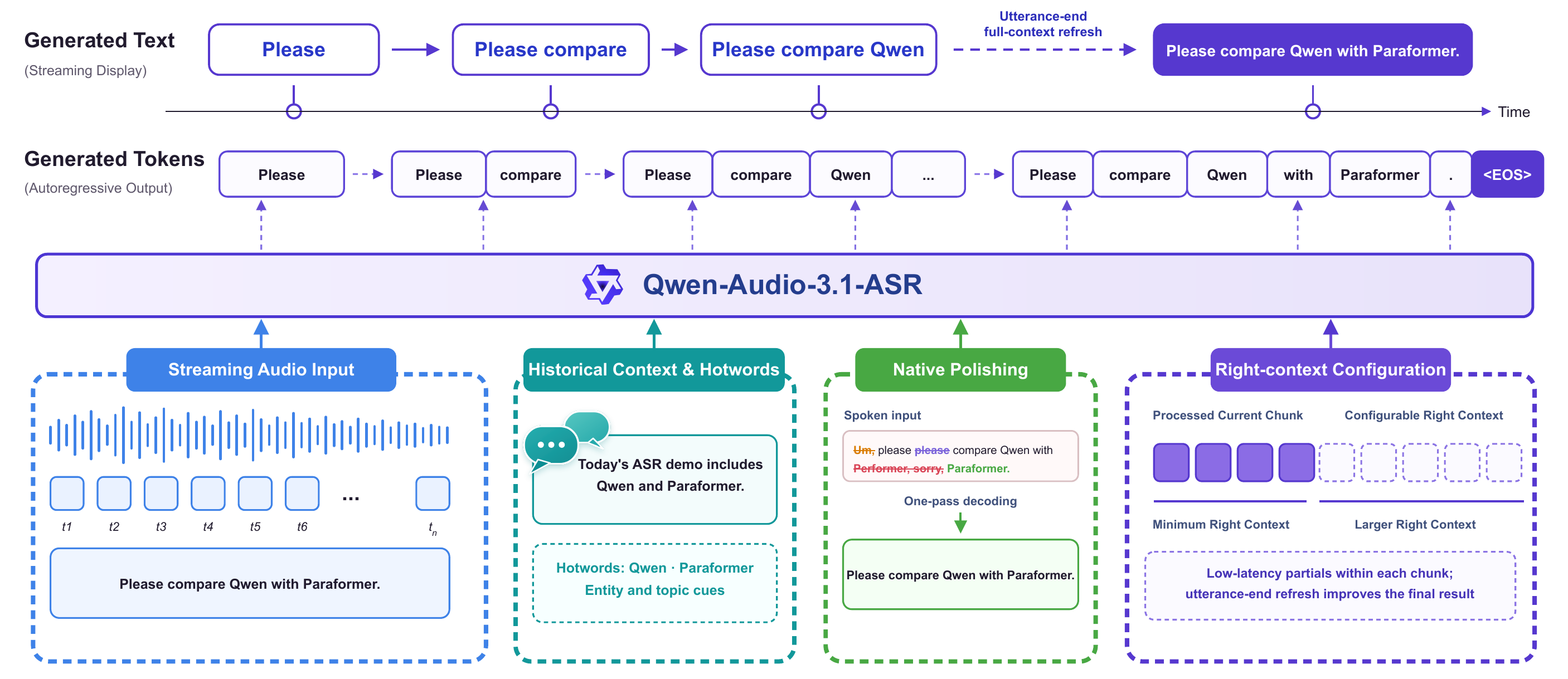}
    \caption{Overview of the Message ASR workflow used in Qwen-Audio-3.1-ASR, following the design of Qwen-Audio-3.0-ASR. Streaming audio is decoded incrementally to provide responsive partial transcripts. Historical context supplies entity and topic cues, native polishing improves readability within the same decoding pass, and configurable right context balances streaming latency and recognition accuracy. At utterance end, the model refreshes the displayed hypothesis using the complete available context.}
    \label{fig:message-asr-overview}
\end{figure*}

The Message ASR deployment profile follows Qwen-Audio-3.0-ASR, retaining its contextual conditioning, hierarchical hotword customization, and native transcription polishing workflow. It is designed for applications such as voice input and conversation documentation, where spontaneous speech must be converted into readable text and terminology may recur across successive utterances.

Historical conversation turns or preceding transcript segments provide topic and entity cues. Hierarchical hotword candidates distinguish a small set of high-confidence terms from a broader candidate pool. These conditions guide recognition together with the acoustic input, supporting the transcription of names, products, and specialized terminology.

A polishing instruction controls whether the output preserves the spoken form or produces a cleaned transcript. When enabled, native polishing removes non-semantic fillers and redundant repetitions, resolves explicit self-corrections, and normalizes punctuation and formatting while preserving the speaker's intended meaning. This processing takes place within recognition rather than through a separate post-recognition rewriting service.

In the interactive workflow, incremental hypotheses provide immediate feedback, and an utterance-end full-context refresh produces the finalized transcript. Language specifications, historical context, hotwords, and polishing can be configured according to the application. These controls remain unchanged from the previous generation and do not imply support for speaker-diarized output or general audio understanding in the Message ASR profile.

\subsection{Speaker-Diarized Transcription}
\label{subsec:deployment-speaker-asr}

The non-streaming Flash and Next variants support speaker-diarized transcription for multi-speaker recordings. Given an audio input and a corresponding task instruction, the model returns recognized text associated with recording-local speaker labels and temporal boundaries.

The output uses the segment format \texttt{[SPK|start-end]text}, where the time interval is expressed in seconds relative to the input recording. Repeated occurrences of a speaker retain the same label within the recording, and intervals from different speakers may overlap. This representation preserves speaker turns, brief responses, and simultaneous speech rather than reducing the conversation to a single unstructured transcript.

The resulting segments support speaker-aware transcript display, navigation to the corresponding audio intervals, and downstream analysis of conversational content. Speaker labels distinguish participants within a recording and do not imply real-world identity recognition.

\subsection{General Audio Understanding}
\label{subsec:deployment-audio-understanding}

The Next variant supports instruction-conditioned understanding of speech, environmental sounds, and music. Given an audio recording together with a task instruction or question, the model produces the corresponding textual response or temporal annotations.

\paragraph{Audio captioning.}
The model generates natural-language descriptions of audible content, including acoustic events, spoken delivery, and musical characteristics. This mode provides information beyond verbatim transcription and is applicable to recordings with or without speech.

\paragraph{Temporal grounding.}
An input query specifies the acoustic content to be localized within the recording. For single-interval localization, the query identifies the start and end boundaries of one target interval, whereas for multi-interval retrieval, it identifies all matching intervals. The resulting annotations provide a unified representation for both temporal grounding tasks.

\paragraph{Audio question answering.}
The model supports both open-ended and multiple-choice questions about the audio. These questions may concern audible content, acoustic characteristics, or temporal relations among events. This interface enables audio-conditioned analysis without first converting the recording into a textual transcript.

These modes complement rather than replace speech recognition. ASR recovers spoken content, speaker-diarized transcription introduces speaker labels and temporal structure, and general audio understanding addresses tasks that require broader acoustic evidence.

\section{Evaluations}
\label{sec:exp}

We evaluate the Qwen-Audio-3.1-ASR family on public benchmarks and internal application-oriented test sets covering conventional ASR, multilingual and dialectal recognition, speaker-diarized transcription, and general audio understanding. We include Qwen-Audio-3.0-ASR results on shared Chinese, English, and multilingual ASR benchmarks to characterize changes across model generations.

For conventional and multilingual ASR, we use the corresponding test sets of AISHELL-1~\citep{bu2017aishell1opensourcemandarinspeech}, AISHELL-2~\citep{du2018aishell2transformingmandarinasr}, FLEURS~\citep{conneau2022fleursfewshotlearningevaluation}, LibriSpeech~\citep{7178964}, WeNetSpeech~\citep{zhang2022wenetspeech10000hoursmultidomain}, GigaSpeechBench~\citep{tu2026gigaspeechbench}, and Common Voice 15~\citep{ardila2020commonvoice}. Public dialect evaluations additionally include KeSpeech and WSYue. Speaker-diarized transcription is evaluated on AISHELL-4, AliMeeting-test, MLC-SLM, and MagicData-RAMC. General audio understanding is evaluated on MMAU~\citep{q31_sakshi2024mmau}, MMSU~\citep{q31_wang2025mmsu}, CochlScene, Nonspeech7k, MELD, NSynth, and DiDiSpeech, with temporal grounding evaluated on Clotho-Moment and DESED. Task-specific subsets, metrics, and evaluation settings are described in the corresponding subsections.

In addition to these public benchmarks, we evaluate the system on internal test sets covering dialect ASR and AST, industry-specific entities and long-tail terminology, hotword customization, transcription polishing, and streaming recognition. These evaluations span diverse acoustic conditions, speaking styles, and application domains. We also present qualitative examples of long-audio contextual recognition to illustrate how preceding transcription history can assist the recognition of recurring entities and technical terms.

\paragraph{Evaluated models and deployment profiles.}
In the result tables and figures, Qwen-Audio-3.1-ASR denotes the base model evaluated through its non-streaming interface for ASR and speaker-diarized transcription. Qwen-Audio-3.1-ASR-Next denotes the Next variant evaluated for speaker-diarized transcription and general audio understanding; where the name Qwen-Audio-3.1-ASR-Flash-Next is used, Flash specifies the model-size designation. Message ASR latency and industry-term recall are evaluated using Qwen-Audio-3.1-ASR-Flash-Message, with streaming outputs used for real-time display and non-streaming outputs used for the utterance-end refresh. Deployment profiles and their capability scope are summarized in Table~\ref{tab:deployment-variants}. Published baseline results and results obtained through our own evaluation are distinguished in the corresponding captions.

\subsection{Public Benchmark Results}
\subsubsection{Chinese and English ASR Benchmarks}

We evaluate Qwen-Audio-3.1-ASR on seven widely used Chinese and English ASR test sets, with results summarized in Table~\ref{tab:asr-os}. Panel A presents baseline results collected from public model repositories and technical reports alongside our results, including Qwen-Audio-3.0-ASR as a reference for the previous generation. Panel B presents results obtained using our unified API-based evaluation pipeline. Given the differences in evaluation settings and result provenance, we discuss the two panels separately and base direct quantitative comparisons primarily on Panel B.

\paragraph{Comparison with published benchmark results.}
The results in Panel A indicate that Qwen-Audio-3.1-ASR maintains broadly comparable Chinese and English recognition performance to Qwen-Audio-3.0-ASR while extending the model family to speaker-diarized transcription and general audio understanding. On the Chinese test sets AISHELL-1, AISHELL-2, and FLEURS-zh, it achieves CERs of 1.08\%, 2.05\%, and 2.25\%, respectively, compared with 1.03\%, 2.02\%, and 2.08\% for the previous generation. On LibriSpeech-clean, the two generations achieve WERs of 1.23\% and 1.19\%, respectively. On LibriSpeech-other, Qwen-Audio-3.1-ASR reduces WER from 2.24\% to 2.19\%, achieving the lowest value in the table. On FLEURS-en, the corresponding WERs are 4.72\% and 4.17\%, representing the largest absolute difference between the two generations among these test sets. Overall, the two generations exhibit similar recognition performance, with differences varying across test sets.

\paragraph{Comparison under the unified evaluation pipeline.}
In Panel B, Qwen-Audio-3.1-ASR achieves the lowest error rate on four of the seven test sets. Compared with the strongest baseline on each of these four test sets, it reduces the error rate by 0.13 percentage points (5.5\% relative) on FLEURS-zh, 0.23 percentage points (4.6\% relative) on FLEURS-en, 0.12 percentage points (8.9\% relative) on LibriSpeech-clean, and 0.32 percentage points (12.7\% relative) on LibriSpeech-other. Fun-ASR-Flash achieves the lowest error rate on the remaining three test sets: AISHELL-1, AISHELL-2, and WeNetSpeech-net. All systems in Panel B are evaluated using the same corrected reference transcripts for WeNetSpeech-net. These results demonstrate competitive recognition performance across Chinese and English benchmarks, with relative performance varying across test sets.

\begin{table*}[t]
\centering
\caption{Performance on \textbf{public ASR benchmark test sets}.
We report CER (\%) for Chinese test sets and WER (\%) for English test sets.
For baselines, Panel A presents results taken from the official public
repositories or the original technical reports, whereas Panel B presents results obtained
using our unified API-based evaluation pipeline on the same test sets. A dash indicates that a result was not reported in the original source. \textbf{Direct quantitative
comparisons are primarily made within Panel B} since the results in Panel A may involve
source-specific preprocessing and scoring conventions. All WeNetSpeech-net results in Panel B use corrected reference transcripts following \href{https://github.com/wenet-e2e/WenetSpeech/issues/63}{WenetSpeech issue~\#63}. All AISHELL-2 results use the test-ios subset.
Lower values indicate better performance, and the best result within each panel
on each test set is bold-faced.}
\label{tab:asr-os}
\scriptsize
\setlength{\tabcolsep}{3.5pt}
\renewcommand{\arraystretch}{1.13}
\begin{tabular*}{\textwidth}{@{\extracolsep{\fill}}llccccccc@{}}
\toprule
\rowcolor{QwenPurpleLight}
\multicolumn{9}{c}{\color{QwenPurple}\bfseries Panel A: Published Benchmark Results} \\
\midrule
Test set & Metric & \shortstack{Kimi\\-Audio} & \shortstack{FireRedASR2\\-LLM} & \shortstack{FireRed\\Audio} & \shortstack{LongCat\\-Next} & \shortstack{Qwen3-ASR\\-1.7B} & \shortstack{Qwen-Audio\\-3.0-ASR} & \shortstack{Qwen-Audio\\-3.1-ASR} \\
\midrule
AISHELL-1 & CER & \textbf{0.60} & 0.64 & 0.71 & 1.47 & -- & 1.03 & 1.08 \\
AISHELL-2 & CER & 2.56 & 2.15 & 2.63 & 2.82 & 2.71 & \textbf{2.02} & 2.05 \\
FLEURS-zh & CER & 2.69 & -- & 3.14 & 3.24 & 2.41 & \textbf{2.08} & 2.25 \\
FLEURS-en & WER & 4.44 & -- & \textbf{2.53} & 5.24 & 3.35 & 4.17 & 4.72 \\
LibriSpeech-clean & WER & 1.28 & -- & \textbf{0.67} & 1.63 & 1.63 & 1.19 & 1.23 \\
LibriSpeech-other & WER & 2.42 & -- & 2.91 & 3.42 & 3.38 & 2.24 & \textbf{2.19} \\
WeNetSpeech-net & CER & 5.37 & 4.44 & 5.18 & 5.98 & 4.97 & \textbf{4.31} & 4.37 \\
\bottomrule
\end{tabular*}
\vspace{8pt}
\begin{tabular*}{\textwidth}{@{\extracolsep{\fill}}llccccccc@{}}
\toprule
\rowcolor{QwenPurpleLight}
\multicolumn{9}{c}{\color{QwenPurple}\bfseries Panel B: Results from Our API Evaluation} \\
\midrule
Test set & Metric & \shortstack{GPT-4o\\Transcribe} & \shortstack{Azure MAI-\\Transcribe-2} & Doubao-ASR & \shortstack{Tencent\\Hy-ASR-3.0-\\preview} & \shortstack{Muse-Voice\\Transcribe-1.0} & \shortstack{Fun-ASR\\-Flash} & \shortstack{Qwen-Audio\\-3.1-ASR} \\
\midrule
AISHELL-1 & CER & 3.34 & 1.49 & 1.54 & 0.78 & 6.28 & \textbf{0.76} & 1.08 \\
AISHELL-2 & CER & 4.47 & 2.82 & 2.66 & 2.15 & 6.80 & \textbf{1.93} & 2.05 \\
FLEURS-zh & CER & 6.14 & 5.38 & 8.18 & 2.38 & 8.34 & 5.52 & \textbf{2.25} \\
FLEURS-en & WER & 5.79 & 5.38 & 6.50 & 5.45 & 7.48 & 4.95 & \textbf{4.72} \\
LibriSpeech-clean & WER & 1.58 & 1.68 & 3.36 & 1.60 & 2.17 & 1.35 & \textbf{1.23} \\
LibriSpeech-other & WER & 3.87 & 3.13 & 4.46 & 2.68 & 5.10 & 2.51 & \textbf{2.19} \\
WeNetSpeech-net & CER & 12.27 & 4.38 & 4.81 & 4.16 & 10.84 & \textbf{3.83} & 4.37 \\
\bottomrule
\end{tabular*}
\end{table*}

\begin{table*}[t]
\centering
\caption{Performance of \textbf{multilingual ASR} on GigaSpeechBench, Common Voice 15, and FLEURS. Results are reported as CER or WER (\%), as specified in the second column; lower is better. Qwen-Audio-3.0-ASR results are taken from its technical report~\citep{qwen_audio_3_0_asr}. Japanese CER for all systems is computed using the same normalization and scoring protocol specified in \href{https://github.com/QwenAudio/Fun-ASR/pull/165}{Fun-ASR PR~\#165}. Macro averages are unweighted means across the listed languages within each benchmark and are reported only when all corresponding results are available. A dash indicates an unavailable result, not necessarily an unsupported language. Bold values denote the lowest error rate in each row.}
\label{tab:multilingual}
\begingroup
\scriptsize
\setlength{\tabcolsep}{4pt}
\renewcommand{\arraystretch}{1.13}
\arrayrulecolor{black}
\begin{adjustbox}{max width=\textwidth}
\begin{tabular}{lccccccccc}
\toprule
Language & Metric & Doubao-ASR & \shortstack{Tencent\\Cloud ASR} & \shortstack{Gemini\\3.1 Pro} & \shortstack{GPT-4o\\Transcribe} & \shortstack{Azure\\MAI-Transcribe-2} & \shortstack{Muse Voice\\Transcribe-1.0} & \shortstack{Qwen-Audio\\-3.0-ASR} & \shortstack{Qwen-Audio\\-3.1-ASR} \\
\midrule

\rowcolor{QwenPurpleLight}
\multicolumn{10}{c}{\color{QwenPurple}\bfseries GigaSpeechBench} \\
Indonesian & WER & 27.73 & 32.55 & 24.47 & 39.52 & 29.18 & 22.90 & 23.68 & \textbf{16.69} \\
Japanese & CER & 29.47 & 34.56 & 39.12 & 45.46 & 34.17 & 32.05 & 28.48 & \textbf{18.47} \\
Thai & CER & 17.65 & 31.40 & 17.37 & 53.55 & 22.42 & 15.27 & 16.22 & \textbf{7.75} \\
Filipino & WER & 25.36 & 33.38 & 37.41 & 37.73 & 32.95 & 25.48 & 29.45 & \textbf{21.69} \\
Vietnamese & WER & 9.53 & 9.61 & 15.55 & 32.89 & 12.77 & 10.76 & 11.54 & \textbf{8.04} \\
Korean & CER & 14.52 & 15.00 & 19.70 & 32.52 & 15.96 & 16.52 & 13.50 & \textbf{7.98} \\
Malay & WER & 35.05 & 41.06 & 40.19 & 51.59 & 38.18 & 38.81 & 34.65 & \textbf{25.04} \\
\cmidrule(lr){1-10}
Macro average & -- & 22.76 & 28.22 & 27.69 & 41.89 & 26.52 & 23.11 & 22.50 & \textbf{15.09} \\
\midrule

\rowcolor{QwenPurpleLight}
\multicolumn{10}{c}{\color{QwenPurple}\bfseries Common Voice 15} \\
Indonesian & WER & 4.25 & 7.79 & 12.92 & 8.74 & 4.33 & 5.98 & 3.66 & \textbf{3.32} \\
Japanese & CER & 6.44 & 12.02 & 11.52 & 11.22 & 7.21 & 12.24 & 5.64 & \textbf{1.40} \\
Thai & CER & 2.32 & 4.17 & 12.33 & 6.64 & 1.90 & 2.94 & 1.56 & \textbf{1.43} \\
Vietnamese & WER & 6.31 & \textbf{5.67} & 10.64 & 15.50 & 5.93 & 7.77 & 8.29 & 8.44 \\
Korean & CER & 3.89 & 8.16 & 4.86 & 5.80 & 2.89 & 7.49 & 3.09 & \textbf{0.73} \\
Spanish & WER & 5.42 & 4.95 & 3.97 & 4.18 & 3.64 & 5.09 & \textbf{3.09} & 3.23 \\
Portuguese & WER & 8.06 & 6.24 & 8.65 & 8.78 & 4.89 & 6.49 & -- & \textbf{4.47} \\
French & WER & 8.89 & 22.06 & 10.51 & 9.26 & 7.38 & 10.72 & \textbf{6.69} & 6.87 \\
\cmidrule(lr){1-10}
Macro average & -- & 5.70 & 8.88 & 9.43 & 8.77 & 4.77 & 7.34 & -- & \textbf{3.74} \\
\midrule

\rowcolor{QwenPurpleLight}
\multicolumn{10}{c}{\color{QwenPurple}\bfseries FLEURS} \\
Indonesian & WER & 5.24 & 12.77 & 3.72 & 3.73 & 2.62 & 8.16 & 2.89 & \textbf{2.01} \\
Japanese & CER & 3.16 & 15.39 & 3.95 & 2.98 & 1.85 & 6.55 & 1.61 & \textbf{0.36} \\
Thai & CER & 6.18 & 11.38 & 6.05 & 4.99 & \textbf{3.85} & 8.27 & 6.73 & 5.37 \\
Filipino & WER & 10.34 & 18.05 & 8.33 & 7.42 & \textbf{5.98} & 9.95 & 11.48 & 10.80 \\
Vietnamese & WER & 6.32 & 15.19 & 4.70 & 3.23 & \textbf{2.42} & 3.97 & 4.95 & 4.46 \\
Korean & CER & 4.86 & 8.93 & 4.24 & 4.06 & 3.92 & 6.87 & 4.94 & \textbf{2.38} \\
Malay & WER & 8.14 & 17.13 & 4.22 & 4.21 & \textbf{4.08} & 9.54 & 8.79 & 5.12 \\
Spanish & WER & 5.31 & 8.09 & 2.62 & 2.13 & 2.04 & 5.61 & \textbf{1.23} & 1.95 \\
Portuguese & WER & 7.19 & 13.71 & 3.46 & 2.80 & 2.53 & 7.12 & \textbf{1.87} & 2.21 \\
\cmidrule(lr){1-10}
Macro average & -- & 6.30 & 13.40 & 4.59 & 3.95 & \textbf{3.25} & 7.34 & 4.94 & 3.85 \\
\bottomrule
\end{tabular}
\end{adjustbox}
\endgroup
\vspace{3pt}
\end{table*}

\subsubsection{Multilingual ASR Benchmarks}

We evaluate the multilingual recognition performance of Qwen-Audio-3.1-ASR on selected language subsets of three public benchmarks: GigaSpeechBench, Common Voice 15, and FLEURS. Table~\ref{tab:multilingual} reports WER or CER according to the language-specific evaluation protocol and includes Qwen-Audio-3.0-ASR as a reference for the previous generation. Macro-average error rates are computed as the unweighted mean across the evaluated languages and are reported only for systems with results on every language in the corresponding benchmark subset.

On the seven East and Southeast Asian languages evaluated in GigaSpeechBench, Qwen-Audio-3.1-ASR achieves the lowest error rate on every language and the lowest macro-average error rate of 15.09\%, compared with 22.50\% for Qwen-Audio-3.0-ASR. Improvements over the previous generation are observed across all seven languages, with particularly large absolute reductions on Japanese, Malay, and Thai.

On the Common Voice 15 subset, which comprises five Asian languages as well as Spanish, Portuguese, and French, Qwen-Audio-3.1-ASR achieves the lowest macro-average error rate of 3.74\% among systems with complete results and the lowest error rate on five of the eight languages: Indonesian, Japanese, Thai, Korean, and Portuguese. Tencent Cloud ASR achieves the lowest WER on Vietnamese, while Qwen-Audio-3.0-ASR retains the lowest WER on Spanish and French. Compared with the previous generation, Qwen-Audio-3.1-ASR shows particularly pronounced improvements on Japanese and Korean, reducing CER from 5.64\% to 1.40\% and from 3.09\% to 0.73\%, respectively. A full-subset macro average is not reported for Qwen-Audio-3.0-ASR because its Portuguese result is unavailable.

On the FLEURS subset, comprising seven Asian languages together with Spanish and Portuguese, Qwen-Audio-3.1-ASR achieves the lowest error rate on Indonesian, Japanese, and Korean. Its macro-average error rate is 3.85\%, compared with 4.94\% for Qwen-Audio-3.0-ASR, with lower error rates on all seven Asian languages. Azure MAI-Transcribe-2 achieves the lowest macro-average error rate of 3.25\% and the lowest error rates on Thai, Filipino, Vietnamese, and Malay, while Qwen-Audio-3.0-ASR retains the lowest WER on Spanish and Portuguese. Overall, these results demonstrate improved multilingual recognition over the previous generation across most shared language subsets, particularly on the evaluated Asian languages, while relative performance varies across languages and benchmarks.

\subsection{Chinese Dialect ASR and AST}
\label{subsec:dialect-results}

We evaluate Qwen-Audio-3.1-ASR against Doubao-ASR and Tencent Hy-ASR-3.0-preview on public dialect benchmarks and internal test sets. ASR requires transcription in the corresponding dialect, whereas AST requires translation of dialectal speech into Mandarin. We report character error rate (CER) for the public benchmarks and internal ASR evaluation, and semantic sentence accuracy for the internal AST evaluation. Macro averages are computed with equal weight for each included subset, separately for each evaluation configuration.

\paragraph{Public dialect benchmarks.}
We evaluate CER on 11 public benchmark subsets: nine KeSpeech subsets under the AST setting and the long- and short-form WSYue Cantonese subsets under the ASR setting. The four WSC Sichuan subsets are excluded from this comparison and its macro averages.

\begin{figure*}[t]
    \centering
    \includegraphics[width=\textwidth]{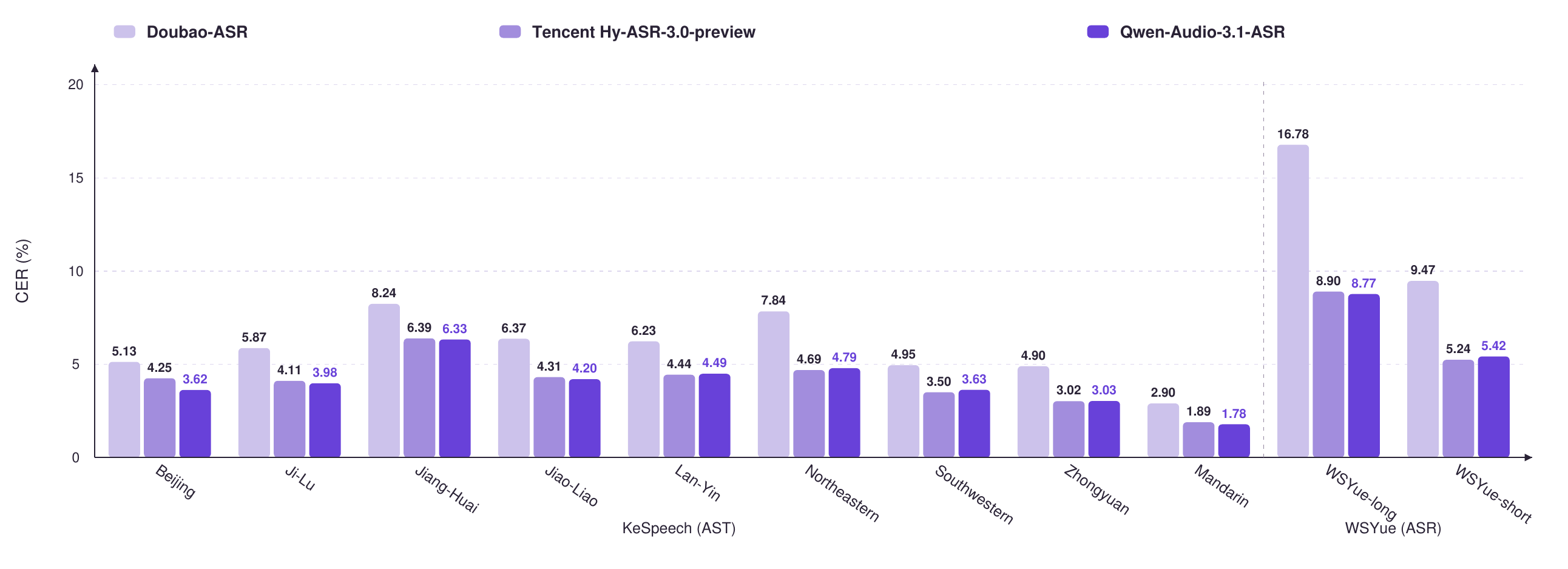}
    \caption{CER (\%) on public dialect ASR and AST benchmarks, comprising nine KeSpeech AST subsets and two WSYue Cantonese ASR subsets. Lower is better.}
    \label{fig:dialect-public-cer}
\end{figure*}

As shown in Figure~\ref{fig:dialect-public-cer}, Qwen-Audio-3.1-ASR achieves a macro-average CER of 4.55\%, compared with 7.15\% for Doubao-ASR and 4.61\% for Tencent Hy-ASR-3.0-preview. It achieves the lowest CER on six subsets: KeSpeech-Beijing, KeSpeech-Ji-Lu, KeSpeech-Jiang-Huai, KeSpeech-Jiao-Liao, KeSpeech-Mandarin, and WSYue-long. These results indicate a substantial improvement over Doubao-ASR and a small macro-average advantage over Tencent Hy-ASR-3.0-preview on the selected public benchmarks.

\paragraph{Internal dialect ASR.}
The internal evaluation includes an ASR configuration covering 16 Chinese dialects and an AST configuration covering 11 dialects. These configurations are evaluated separately rather than assigning each dialect exclusively to one task. For the internal ASR evaluation, all subsets require transcription in the corresponding dialect and are evaluated using CER.

\begin{figure*}[t]
    \centering
    \includegraphics[width=\textwidth]{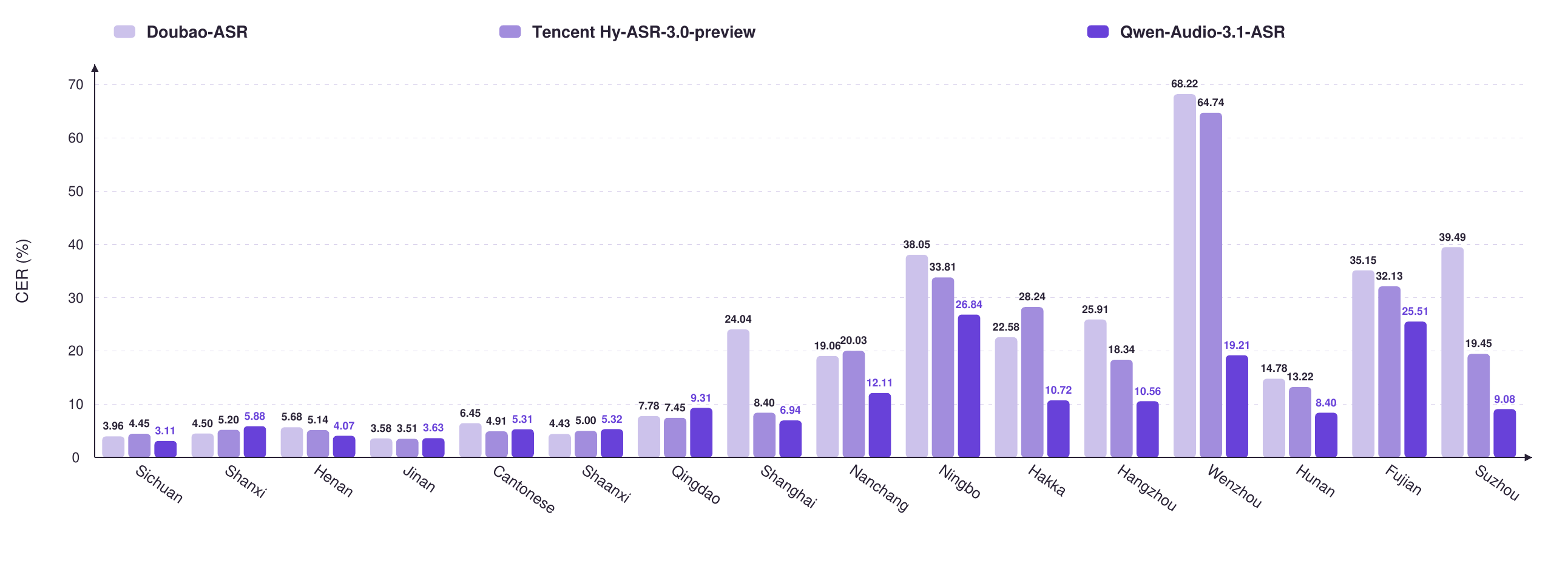}
    \caption{CER (\%) on the internal ASR test set covering 16 Chinese dialects. All subsets use the ASR setting. Doubao-ASR does not explicitly list Henan, Qingdao, Nanchang, Ningbo, Hangzhou, Wenzhou, Hunan, or Suzhou in its publicly documented dialect coverage. Tencent Hy-ASR-3.0-preview does not explicitly list Shanxi, Shanghai, Ningbo, Hakka, Hangzhou, Wenzhou, Fujian, or Suzhou. Absence from these lists does not necessarily imply lack of recognition capability. Lower is better.}
    \label{fig:dialect-cer}
\end{figure*}

As shown in Figure~\ref{fig:dialect-cer}, Qwen-Audio-3.1-ASR achieves the lowest CER on 11 of the 16 internal ASR subsets. Its macro-average CER is 10.38\%, compared with 20.23\% for Doubao-ASR and 17.13\% for Tencent Hy-ASR-3.0-preview, corresponding to absolute reductions of 9.85 and 6.75 percentage points, respectively. The largest improvement occurs on Wenzhou, where CER decreases to 19.21\%, compared with 68.22\% and 64.74\% for the two baselines. Qwen-Audio-3.1-ASR also achieves the lowest CER on Shanghai, Nanchang, Sichuan, Ningbo, Hakka, Hangzhou, Henan, Hunan, Fujian, and Suzhou.

\paragraph{Internal dialect AST.}
For the internal AST evaluation, we report \emph{semantic sentence accuracy}, defined as the percentage of outputs that preserve the reference meaning without errors in named entities or other key information. Qwen3.7-Max evaluates each model output independently against the reference text. To validate this automated evaluation, we manually annotate a subset of the evaluation data using the same criterion, achieving 95.35\% agreement between human and Qwen3.7-Max judgments. Based on this high agreement, we use Qwen3.7-Max to score the full evaluation set.

An output is considered semantically incorrect if it contains an incorrect named entity, changes or omits key information such as a participant, action, object, time, location, quantity, negation, causal relation, or fact, or introduces punctuation or text-normalization errors that alter the meaning. Minor wording, grammatical, punctuation, or formatting differences are accepted when the core meaning and key information remain unchanged. Semantically equivalent dialectal and Mandarin expressions are also accepted: this metric assesses semantic fidelity, independently of whether the output conforms to the requested language form.

Semantic sentence accuracy is computed as:
\begin{equation}
    \mathrm{Semantic\ Sentence\ Accuracy}(\%)
    = \frac{100}{N}\sum_{i=1}^{N}\mathbf{1}[z_i=0].
\end{equation}

Here, $N$ is the number of evaluated utterances, $z_i=1$ indicates that the $i$-th output contains a semantic error under the criteria above, and $z_i=0$ otherwise. The indicator function $\mathbf{1}[\cdot]$ equals 1 when its condition holds and 0 otherwise.

\begin{figure*}[t]
    \centering
    \includegraphics[width=\textwidth]{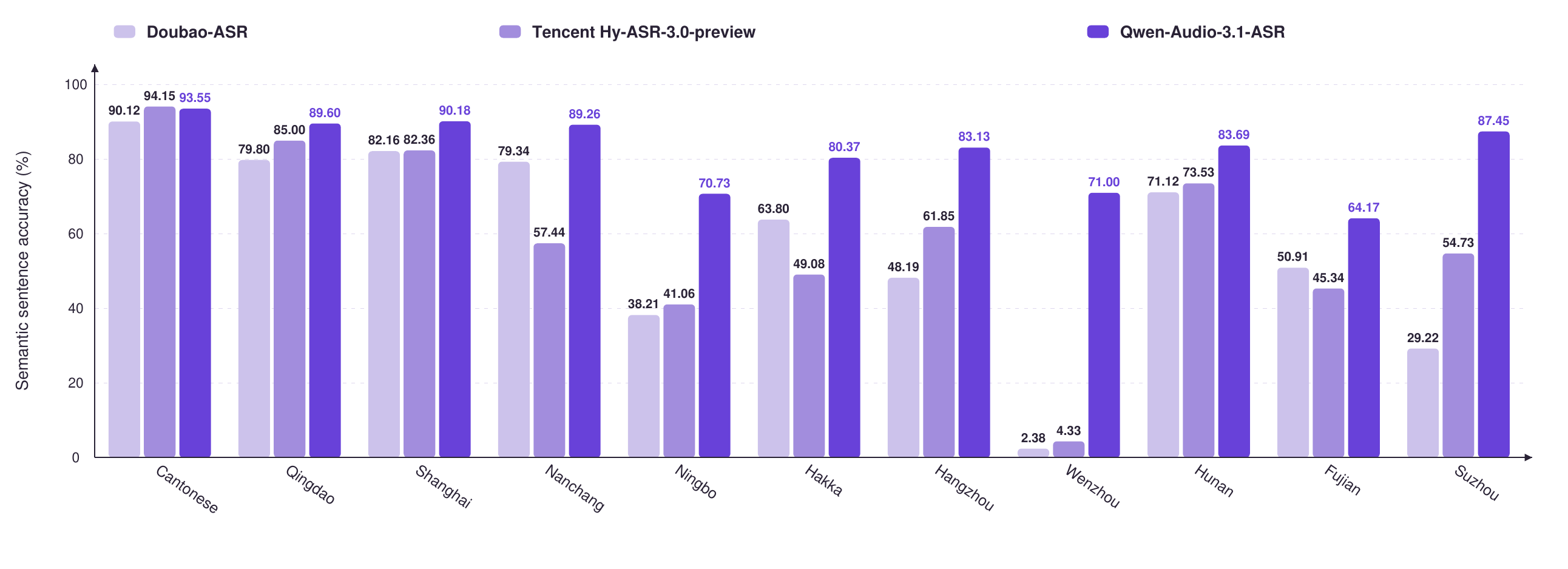}
    \caption{Semantic sentence accuracy (\%) on the internal AST test set covering 11 Chinese dialects, with Mandarin translation as the target task. Qwen3.7-Max assesses whether each output preserves the reference meaning, including named entities and key information. Semantically equivalent expressions and minor differences that do not affect meaning are accepted. Higher is better.}
    \label{fig:dialect-consistency}
\end{figure*}

On the internal AST test set (Figure~\ref{fig:dialect-consistency}), we compare Qwen-Audio-3.1-ASR with each baseline on subsets corresponding to that baseline's explicitly documented dialect coverage. Against Doubao-ASR, the comparison includes Shanghai, Hakka, and Cantonese. Qwen-Audio-3.1-ASR achieves higher semantic sentence accuracy on all three subsets, with a macro-average accuracy of 88.03\%, compared with 78.69\% for Doubao-ASR, an improvement of 9.34 percentage points. Against Tencent Hy-ASR-3.0-preview, the comparison includes Nanchang, Hunan, Cantonese, and Qingdao. Qwen-Audio-3.1-ASR achieves a macro-average accuracy of 89.03\%, compared with 77.53\% for Tencent Hy-ASR-3.0-preview, and performs better on three of the four subsets. The largest gain is observed on Nanchang, where accuracy increases from 57.44\% to 89.26\%. On Cantonese, Qwen-Audio-3.1-ASR achieves 93.55\%, slightly below Tencent Hy-ASR-3.0-preview at 94.15\%. These pairwise comparisons indicate competitive semantic preservation under our dialect-to-Mandarin AST evaluation setting.

\begin{figure}[t]
    \centering
    \includegraphics[width=\linewidth]{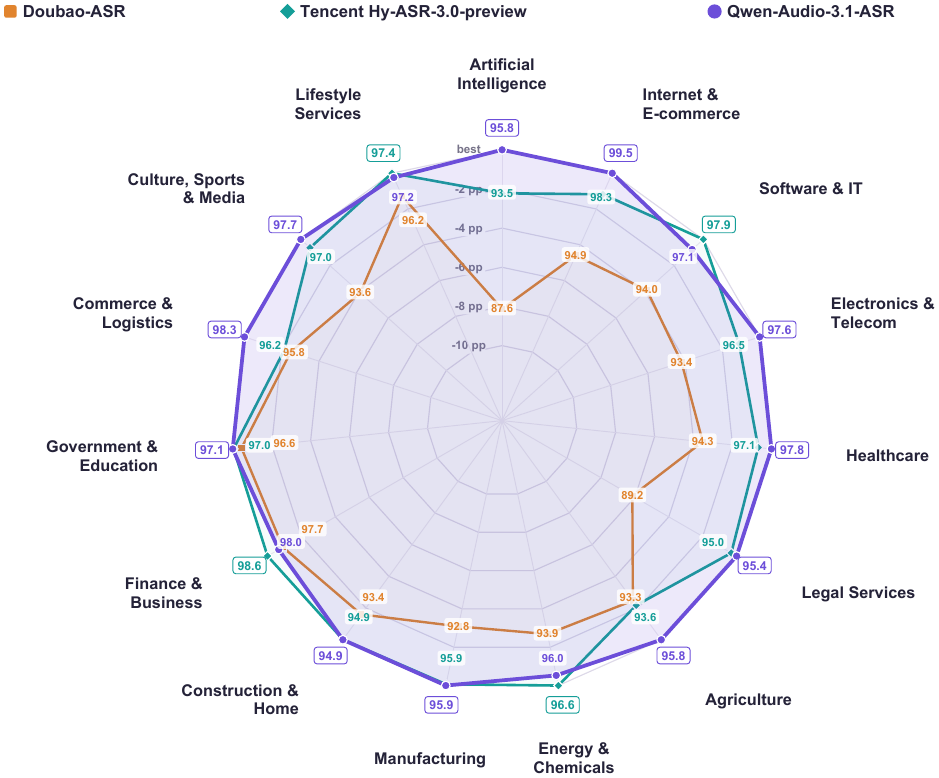}
    \caption{Entity recall (\%) across 15 industry domains. Radial positions represent percentage-point gaps from the highest recall in each domain, with the outer ring indicating the domain-specific best result. Numerical labels show the original recall values. Higher recall indicates better performance.}
    \label{fig:entity-recall}
\end{figure}

\subsection{Industry-Domain Entity Recognition}
\label{subsec:industry-terms}

Reliable recognition of domain-specific terminology is essential for practical ASR applications. Financial and medical terms, software libraries, API names, organization names, and other specialized entities may be underrepresented in general-purpose training data, making them susceptible to substitution and omission errors. To improve coverage of these long-tail entities, we continually mine domain-specific vocabularies and construct corresponding ASR training examples using the entity-mining and data-synthesis pipeline described in Section~\ref{subsec:SFT}.

Figure~\ref{fig:entity-recall} compares Doubao-ASR, Tencent Hy-ASR-3.0-preview, and Qwen-Audio-3.1-ASR on an internal entity-recognition evaluation set spanning 15 industry domains. Qwen-Audio-3.1-ASR achieves the highest entity recall in nine domains and ties for the highest recall in Construction \& Home and Manufacturing. Its recall ranges from 94.9\% to 99.5\% across the evaluated domains. The largest improvements over the stronger competing system are observed in Artificial Intelligence, Agriculture, and Commerce \& Logistics, with gains of 2.3, 2.2, and 2.1 percentage points, respectively. Tencent Hy-ASR-3.0-preview retains higher recall in Software \& IT, Energy \& Chemicals, Finance \& Business, and Lifestyle Services. Overall, these results indicate strong recognition of industry-specific entities across a diverse range of domains. The observed performance is consistent with the intended benefits of the entity-mining and data-synthesis pipeline; however, this comparison does not isolate its contribution from other improvements incorporated into Qwen-Audio-3.1-ASR.

\subsection{Hotword Customization}
\label{subsec:hotword}

Qwen-Audio-3.1-ASR incorporates a hierarchical hotword customization mechanism that distinguishes between two priority tiers: a high-priority set (\textbf{P0}) and a broader candidate set (\textbf{P1}). During decoding, P0 candidates receive stronger conditioning than P1 candidates. This priority-aware design emphasizes high-confidence terms while retaining coverage of a broader set of potentially relevant entities.
Table~\ref{tab:hotword-recall} compares hotword recall with and without hotword conditioning across six categories of long-tail entities.
With conditioning enabled, Qwen-Audio-3.1-ASR achieves the highest recall among the evaluated systems in every P0 category, with recall ranging from 97.95\% to 99.46\%. These results demonstrate the effectiveness of the proposed mechanism in recognizing high-priority long-tail entities.
For subject terms and trending buzzwords, hotword conditioning also improves P1 recall, although the gains are smaller than those for the corresponding P0 categories. This pattern is consistent with the weaker conditioning assigned to P1 candidates. However, Qwen-Audio-3.1-ASR does not achieve the highest conditioned recall in these two P1 categories, highlighting the distinction between prioritizing high-confidence terms and maximizing recall over a broader candidate set.

\begin{table*}[t]
\centering
\caption{Hotword recall (\%) by entity category and priority tier.
Results are reported without (w/o) and with (w/) hotword conditioning.
P0 denotes high-priority hotwords, and P1 denotes the broader candidate set.
Higher values indicate better performance. Bold values denote the highest
recall among systems with hotword conditioning in each row.}
\label{tab:hotword-recall}
\begingroup
\small
\setlength{\tabcolsep}{4pt}
\renewcommand{\arraystretch}{1.15}
\begin{tabular*}{\textwidth}{@{\extracolsep{\fill}}lc*{6}{r}@{}}
\toprule
\multicolumn{2}{c}{}
& \multicolumn{2}{c}{Doubao-ASR}
& \multicolumn{2}{c}{Fun-ASR-Flash}
& \multicolumn{2}{c}{Qwen-Audio-3.1-ASR} \\
\cmidrule(lr){3-4}\cmidrule(lr){5-6}\cmidrule(lr){7-8}
Category & Tier & w/o & w/ & w/o & w/ & w/o & w/ \\
\midrule
Person names       & P0 & 32.27 & 82.64 & 37.33 & 95.26 & 62.12 & \textbf{99.43} \\
\addlinespace[3pt]
Subject terms      & P0 & 75.14 & 93.06 & 71.84 & 92.53 & 69.36 & \textbf{99.42} \\
                   & P1 & 84.36 & 93.04 & 86.02 & \textbf{96.73} & 88.46 & 95.31 \\
\addlinespace[3pt]
Trending buzzwords & P0 & 48.24 & 91.95 & 43.26 & 82.53 & 63.08 & \textbf{99.46} \\
                   & P1 & 51.24 & \textbf{89.48} & 53.92 & 87.00 & 74.00 & 88.34 \\
\addlinespace[3pt]
AI entities        & P0 & 44.21 & 65.24 & 79.88 & 90.55 & 88.22 & \textbf{99.39} \\
Brand names        & P0 & 44.18 & 57.88 & 80.14 & 95.55 & 83.83 & \textbf{97.95} \\
Product entities   & P0 & 67.59 & 78.70 & 73.64 & 89.92 & 72.87 & \textbf{99.07} \\
\bottomrule
\end{tabular*}
\endgroup
\end{table*}

\subsection{Long-Audio Context Modeling}
\label{subsec:long-audio-context}

Accurate transcription may require contextual information beyond the current audio segment, particularly when resolving homophones, technical terminology, and entity names. For example, a technical filename may be confused with a phonetically similar common expression, while a person's name may be difficult to identify from an isolated segment. Independently decoding each segment limits the use of preceding conversational information that could help disambiguate such cases.

Building on the contextual supervised fine-tuning described in Section~\ref{subsec:SFT}, Qwen-Audio-3.1-ASR conditions the transcription of each audio segment on recently recognized text from the same session. This transcription history provides a dynamically updated source of context, enabling the model to incorporate information from preceding segments without requiring a manually maintained hotword list.

Table~\ref{tab:long-context-examples} presents representative examples of context-dependent corrections. Previously mentioned entities and technical terms, including codebase names, abbreviations, and personal names, provide contextual cues for recognizing subsequent occurrences. In these examples, conditioning on the transcription history improves recognition consistency across segments, whereas decoding the same audio without prior context produces phonetically plausible but incorrect alternatives.

\begin{table*}[t]
\centering

\begin{CJK*}{UTF8}{gbsn}

\caption{
Representative corrections enabled by long-audio context.
The context cues occurred earlier in the same session.
Colored spans highlight recognition errors and their
context-conditioned corrections.
}
\label{tab:long-context-examples}

\begingroup
\small
\renewcommand{\arraystretch}{1.22}
\setlength{\tabcolsep}{5pt}
\arrayrulecolor{QwenPurple!35}

\begin{tabular}{
  @{}
  V{0.27\textwidth}
  V{0.23\textwidth}
  V{0.22\textwidth}
  V{0.22\textwidth}
  @{}
}
\toprule

\rowcolor{QwenPurple!14}
\textbf{Earlier context cue}
&
\textbf{Reference}
&
\textbf{Without long context}
&
\textbf{With long context}
\\

\midrule

\texttt{xsltproc style.xsl input.xml}
occurred earlier
&
\texttt{style.xsl data.xml > out.html}
&
\texttt{\bad{xsl} data.xml > out.html}
&
\texttt{\good{style.xsl} data.xml > out.html}
\\

\rowcolor{QwenLavender}
Earlier explanation:
“这叫法条竞合”
&
法条竞合
&
\bad{划掉即可}
&
\good{法条竞合}
\\

The term \texttt{unigram}
occurred twice earlier
&
我们称之为 \texttt{unigram}
&
我们称之为 \bad{\texttt{unique}}
&
我们称之为 \good{\texttt{unigram}}
\\

\rowcolor{QwenLavender}
\texttt{TCC} was repeatedly
established as the metric name
&
\texttt{our TCC will be...}
&
\texttt{our \bad{GGC} will be...}
&
\texttt{our \good{TCC} will be...}
\\

The name “洛军是阿占的儿子”
occurred earlier
&
洛军是阿占的儿子
&
\bad{洛君}是阿占的儿子
&
\good{洛军}是阿占的儿子
\\

\rowcolor{QwenLavender}
The technical term
\texttt{softmax} occurred earlier
&
\texttt{softmax} 形式
&
\texttt{\bad{softmap}} 形式
&
\texttt{\good{softmax}} 形式
\\

The mode name
\texttt{debug} occurred earlier
&
只能 \texttt{debug} 模式
&
只能 \bad{第八个} 模式
&
只能 \good{\texttt{debug}} 模式
\\

\rowcolor{QwenLavender}
The name “张贵生”
occurred earlier
&
张贵生
&
\bad{张桂生}
&
\good{张贵生}
\\

\bottomrule
\end{tabular}

\endgroup
\end{CJK*}
\end{table*}

\begin{figure}[h]
    \centering
    \includegraphics[width=1\linewidth]{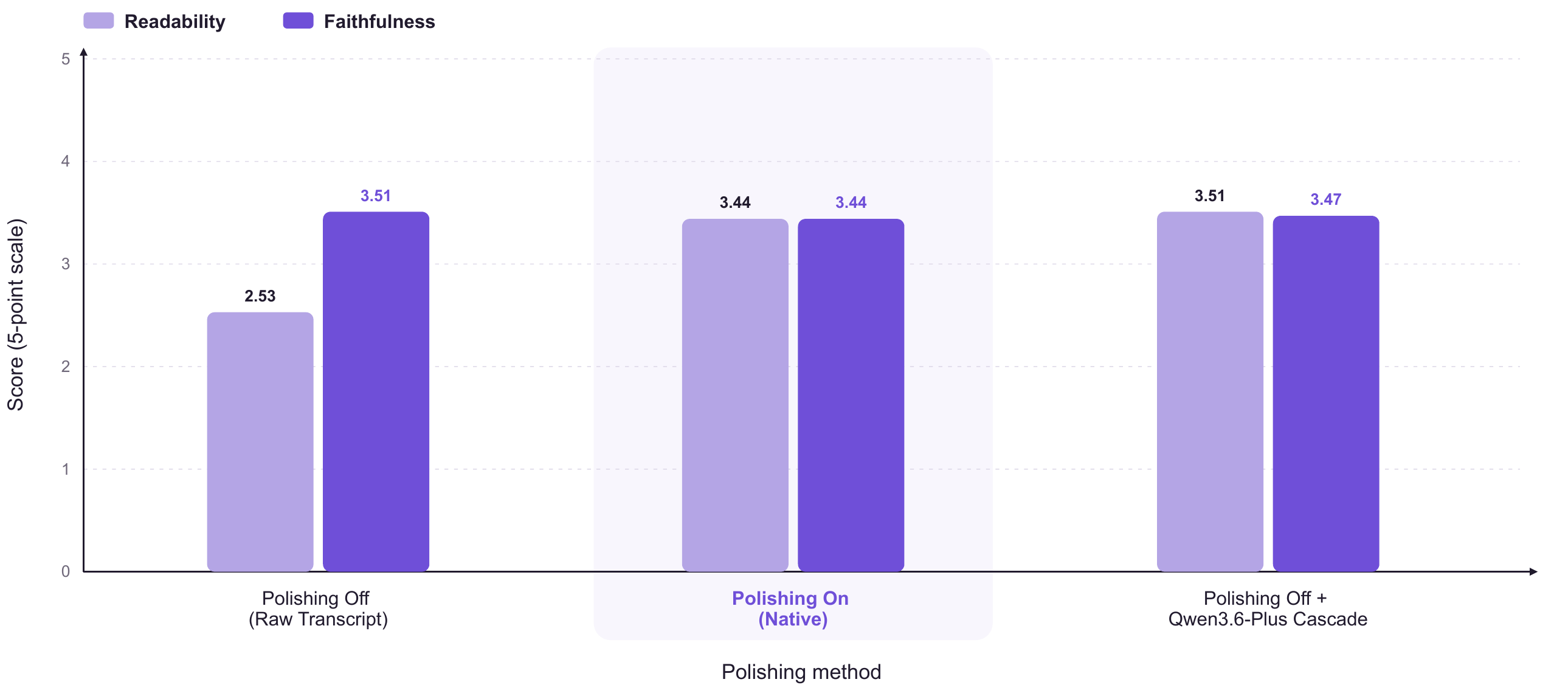}
    \caption{Readability and faithfulness of native polishing vs.\ the raw transcript and vs.\ an external ASR+LLM cascade.}
    \label{fig:polishing}
\end{figure}

\subsection{Native Transcription Polishing}
\label{subsec:native-polishing}

Conventional transcription pipelines typically handle fillers, disfluent repetitions, and self-corrections through a separate text-refinement stage, in which an LLM rewrites the ASR output. Although this approach separates recognition from transcript refinement, it requires an additional model invocation and introduces a sequential processing dependency before the polished transcript becomes available.

Qwen-Audio-3.1-ASR integrates transcription polishing directly into the recognition process. An instruction flag selects between raw transcription and polished output within a single decoding pass. When polishing is enabled, the model removes fillers and disfluent repetitions, incorporates the speaker's final stated corrections, and standardizes punctuation and formatting without an additional model call. Figure~\ref{fig:polishing} compares native polishing with raw transcription and a two-stage cascade in which the same raw ASR output is refined by Qwen3.6-Plus. On an internal five-point rating scale, native polishing improves readability from 2.53 to 3.44, compared with 3.51 for the cascade. Its faithfulness score is 3.44, close to the cascade's 3.47 and slightly below the raw transcript's 3.51. These results indicate that native polishing provides a substantial improvement in readability with a modest reduction in rated faithfulness, while eliminating the separate text-refinement invocation.

\begin{figure*}[t]
    \centering
    \includegraphics[width=\textwidth]{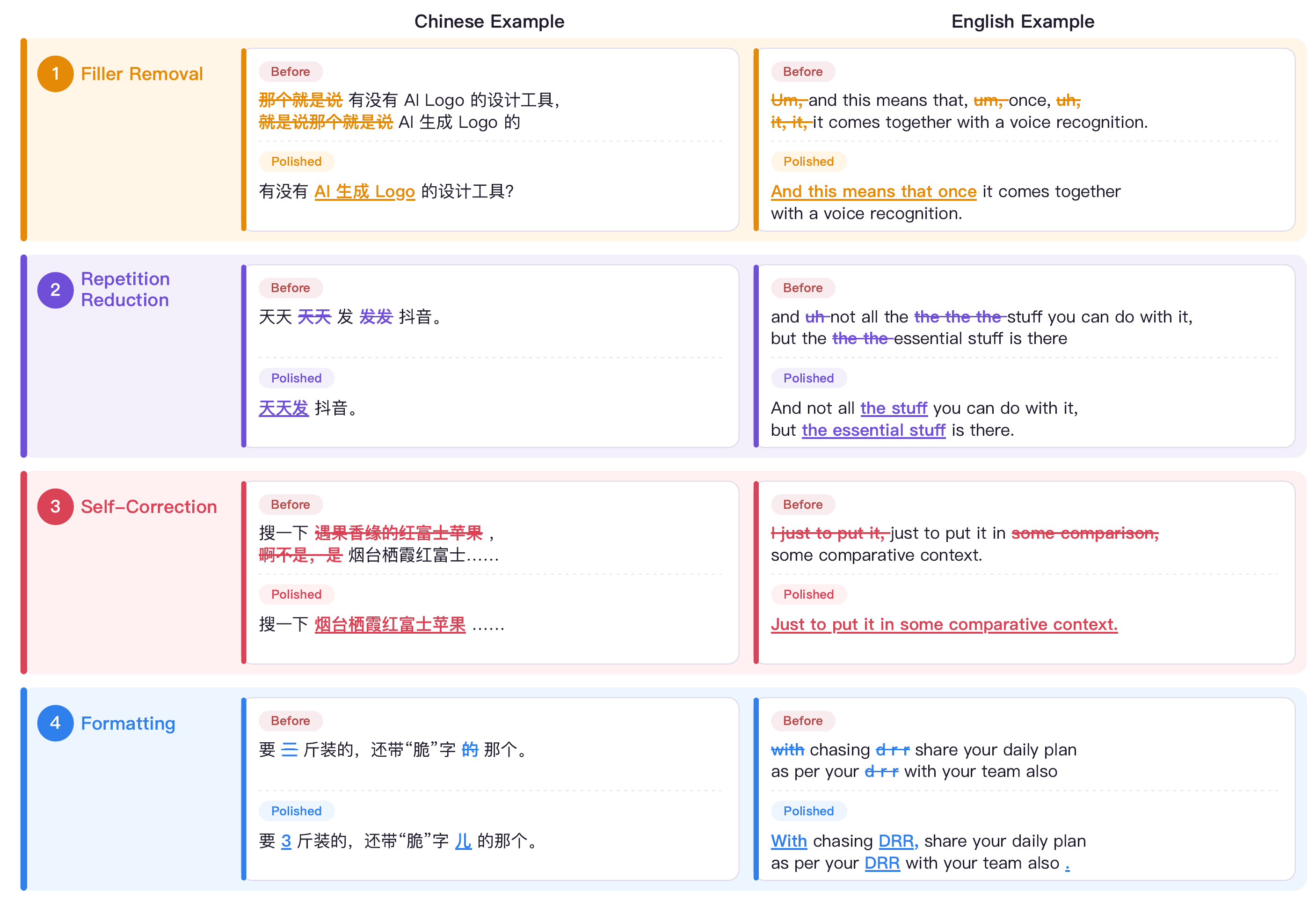}
    \caption{Representative Chinese and English examples of native transcription polishing. Qwen-Audio-3.1-ASR removes fillers and disfluent repetitions, incorporates self-corrections, and standardizes sentence structure, punctuation, capitalization, and acronyms. The examples illustrate improvements in readability while largely retaining the original meaning.}
    \label{fig:bilingual-polishing-examples}
\end{figure*}

Figure~\ref{fig:bilingual-polishing-examples} presents representative examples in Chinese and English. In the English examples, the model removes hesitation fillers such as \textit{um} and \textit{uh}, consolidates repeated words such as \textit{the the the}, and reformulates false starts into more coherent expressions. For instance, the fragmented phrase \textit{I just to put it / some comparison} is rendered as \textit{Just to put it in some comparative context}. The model also standardizes capitalization, punctuation, and acronym formatting, converting the spoken letter sequence \textit{d r r} into \textit{DRR}. These examples illustrate how native polishing improves transcript fluency and readability while largely retaining the meaning conveyed in the original speech.

\subsection{Streaming Display and Non-streaming Refresh in Message ASR}
\label{subsec:industry-terms_message}

Message ASR combines streaming recognition for real-time transcript display with non-streaming recognition for refreshing the transcript at utterance end. We compare qwen-audio-3.1-asr-flash-message with Doubao Streaming Speech Recognition Model 2.0, abbreviated as Qwen Message and Doubao, respectively. We evaluate streaming latency and industry-term recall separately for the streaming and non-streaming outputs to distinguish the responsiveness and quality of the intermediate displayed text from the quality of the final refreshed transcript.

\begin{table}[t]
    \centering
    \caption{Streaming latency of Doubao Streaming ASR 2.0 and Qwen-Audio-3.1-ASR-Flash-Message. All values are in milliseconds; lower is better.}
    \label{tab:message-asr-latency}
    \begingroup
    \small
    \renewcommand{\arraystretch}{1.2}
    \setlength{\tabcolsep}{8pt}
    \arrayrulecolor{black}
    \begin{tabular}{lrr}
        \toprule
        Metric & Doubao Streaming ASR 2.0 & Qwen-Audio-3.1-ASR-Flash-Message \\
        \midrule
        Text-emission latency   & 530 & \textbf{220} \\
        First-character latency & 570 & \textbf{100} \\
        Last-character latency  & 307 & \textbf{103} \\
        \bottomrule
    \end{tabular}
    \endgroup
\end{table}

\begin{figure*}[t]
    \centering
    \includegraphics[width=\textwidth]{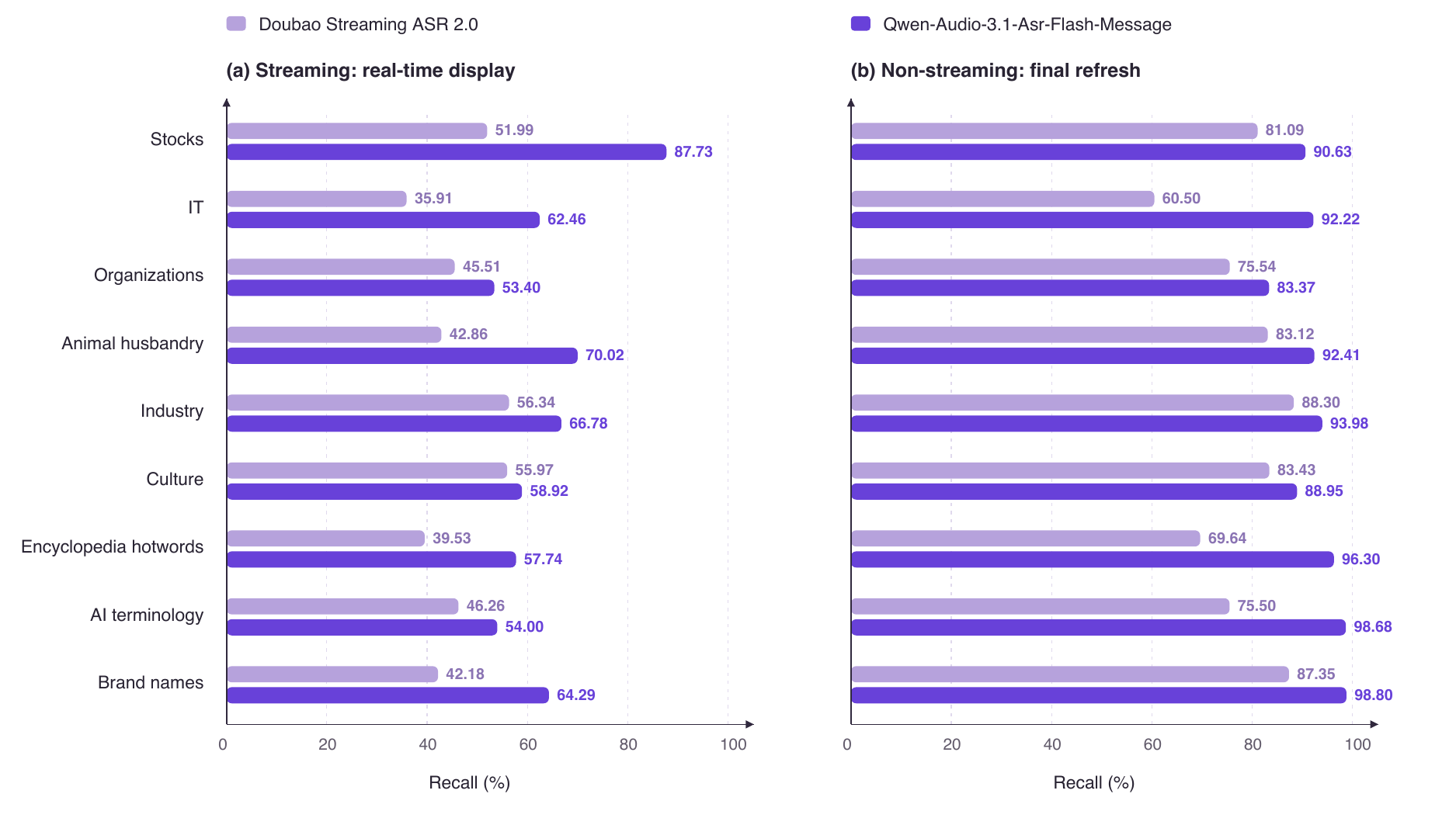}
    \caption{Industry-term recall on nine domain-specific test sets for (a) streaming outputs used for real-time display and (b) non-streaming outputs used for the final transcript refresh. Doubao Streaming ASR 2.0 abbreviates Doubao Streaming Speech Recognition Model 2.0. Each panel compares the two systems within the same output stage. Recall is reported as a percentage; higher is better.}
    \label{fig:message-asr-industry-recall}
\end{figure*}

We measure display latency as the elapsed time between the end of the corresponding spoken content and its appearance in the displayed transcript. Doubao Streaming ASR 2.0 is evaluated through its API, while Qwen-Audio-3.1-ASR-Flash-Message is evaluated using a self-hosted service running on NVIDIA H100 GPUs. Both systems are tested with 10 concurrent audio streams. Network latency is assumed to be comparable between the two systems.
As shown in Table~\ref{tab:message-asr-latency}, Qwen-Audio-3.1-ASR-Flash-Message achieves lower latency on all three reported metrics. Its first-character latency is 100~ms, compared with 570~ms for Doubao Streaming ASR 2.0, while text-emission latency decreases from 530~ms to 220~ms and last-character latency from 307~ms to 103~ms. These results indicate faster transcript display under the evaluated conditions, before the non-streaming refresh at utterance end.

For streaming recognition, Figure~\ref{fig:message-asr-industry-recall}(a) shows that Qwen-Audio-3.1-ASR-Flash-Message achieves higher industry-term recall than Doubao Streaming ASR 2.0 on all nine domain-specific test sets. Its unweighted mean recall is 63.93\%, compared with 46.28\% for Doubao Streaming ASR 2.0, a difference of 17.64 percentage points. The largest gains occur in stocks (+35.74 points), animal husbandry (+27.16 points), and IT (+26.55 points), indicating improved recognition of domain-specific terms in the intermediate displayed transcripts.

For the non-streaming refresh, Figure~\ref{fig:message-asr-industry-recall}(b) shows that Qwen-Audio-3.1-ASR-Flash-Message again achieves higher industry-term recall than Doubao Streaming ASR 2.0 on all nine sets, with an unweighted mean of 92.82\%, compared with 78.27\%, a difference of 14.54 percentage points. The largest gains occur in IT (+31.72 points), encyclopedia hotwords (+26.66 points), and AI terminology (+23.18 points). Both systems achieve higher recall after the non-streaming refresh than in their corresponding streaming outputs on every evaluated domain. Together, these results illustrate the complementary roles of the two stages: streaming recognition provides responsive intermediate text, while non-streaming recognition improves industry-term recall in the final refreshed transcript.

\subsection{Speaker-diarized ASR}
\label{subsec:SDR}

We evaluate joint speaker diarization and transcription on four test sets using DER and cpWER. Table~\ref{tab:speaker-diarized-asr} compares the two Qwen-Audio-3.1-ASR variants with a cascaded Fun-ASR baseline and three other systems.

\begin{table*}[t]
    \centering
    \caption{Speaker-diarized ASR results on four evaluation sets. To ensure an equitable comparison and mitigate inference instability common in LALMs during long-form audio processing, all test sets are standardized into approximately 4-minute audio chunks. The chunking methodology and evaluation scripts are provided in the \href{https://github.com/QwenAudio/Fun-ASR/tree/main/tools/sdr_eval_demo/}{SDR Evaluation Toolkit}. Fun-ASR is evaluated as a cascaded system. DER, WER, and cpWER are reported as percentages; lower is better. The best result in each column is shown in bold. Systems marked with an asterisk (*) use results taken directly from their respective technical reports.}
    \label{tab:speaker-diarized-asr}
    \small
    \setlength{\tabcolsep}{4.5pt}
    \begingroup
    \renewcommand{\arraystretch}{1.2}
    \arrayrulecolor{black}
    \resizebox{\textwidth}{!}{
    \begin{tabular}{lrrrrrrrrrrrr}
        \toprule
        & \multicolumn{3}{c}{AISHELL-4}
        & \multicolumn{3}{c}{AliMeeting-test}
        & \multicolumn{3}{c}{MLC-SLM}
        & \multicolumn{3}{c}{MagicData-RAMC} \\
        \cmidrule(lr){2-4}\cmidrule(lr){5-7}\cmidrule(lr){8-10}\cmidrule(lr){11-13}
        System & DER & WER & cpWER & DER & WER & cpWER & DER & WER & cpWER & DER & WER & cpWER \\
        \midrule
        MOSS-Transcribe-Diarize (MOSI) & 4.20 & 14.33 & 15.23 & 3.38 & 19.47 & 17.23 & 3.20 & 11.71 & 14.06 & 4.79 & 10.88 & 14.00 \\
        Muse-Voice-Transcribe-1.0 (Meta) & 3.04 & 17.78 & 19.04 & 6.16 & 26.93 & 27.97 & 3.48 & 12.64 & 14.60 & 4.55 & 12.55 & 15.47 \\
        Gemini-3.5-Transcribe (Google) & 13.99 & 24.38 & 36.39 & 21.84 & 33.18 & 45.16 & 12.71 & 14.47 & 23.05 & 11.44 & 15.29 & 25.32 \\
        Mai-Transcribe-2 (Microsoft) & 5.96 & 14.92 & 22.25 & 8.52 & 23.83 & 27.65 & 7.01 & 12.26 & 15.63 & 5.42 & 11.99 & 14.98 \\
        Fun-ASR-API (Cascaded) & 8.45 & 17.08 & 21.22 & 9.03 & 25.37 & 34.68 & 6.17 & 29.26 & 33.83 & 5.37 & 11.52 & 20.42 \\
        VibeVoice-ASR* (Microsoft) & - & - & 24.99 & - & - & 29.33 & - & - & 14.81 & - & - & - \\
        SoulX-Transcriber* (Soul) & - & 14.49 & 17.82 & - & 16.22 & 16.99 & - & - & - & - & - & - \\
        \midrule
        \rowcolor{QwenPurpleLight}
        Qwen-Audio-3.1-ASR & 3.26 & 13.02 & 15.02 & \textbf{2.11} & 17.23 & 15.78 & 1.80 & \textbf{9.89} & \textbf{12.35} & \textbf{3.08} & \textbf{10.21} & \textbf{13.32} \\
        \rowcolor{QwenPurpleLight}
        Qwen-Audio-3.1-ASR-Next & \textbf{2.70} & \textbf{12.03} & \textbf{13.51} & 2.15 & \textbf{13.89} & \textbf{12.04} & \textbf{1.78} & 10.32 & 12.60 & 3.18 & 10.50 & 13.58 \\
        \bottomrule
    \end{tabular}}
    \endgroup
\end{table*}

The two Qwen-Audio-3.1-ASR variants collectively achieve the lowest error rates across all 12 dataset--metric combinations, covering DER, WER, and cpWER on four test sets. Qwen-Audio-3.1-ASR-Next achieves the lowest DER on AISHELL-4 and MLC-SLM, as well as the lowest WER and cpWER on AISHELL-4 and AliMeeting-test. Qwen-Audio-3.1-ASR achieves the lowest DER on AliMeeting-test, the lowest WER and cpWER on MLC-SLM, and the lowest values for all three metrics on MagicData-RAMC. The largest cpWER difference between the two variants occurs on AliMeeting-test, where Next achieves 12.04\%, compared with 15.78\% for Qwen-Audio-3.1-ASR, an absolute reduction of 3.74 percentage points. These results highlight complementary strengths across speaker diarization and transcription.

Both variants outperform the cascaded Fun-ASR baseline on all four test sets in both DER and cpWER. Next also achieves lower error rates than all non-Qwen systems across all eight metrics. These results demonstrate strong speaker diarization and transcription performance across the evaluated test sets, with the relative strengths of the two Qwen variants depending on the dataset and metric.

\subsection{General Audio Understanding}
\label{sec:audio-understanding}

We evaluate Qwen-Audio-3.1-ASR-Flash-Next on seven comprehensive and specialized audio understanding benchmarks measured by accuracy, and two temporal grounding benchmarks measured by mean Intersection over Union~(mIoU). Table~\ref{tab:audio-understanding} compares its results with three competing systems.

\begin{table*}[t]
    \centering
    \caption{General audio understanding and temporal grounding results for Qwen-Audio-3.1-ASR-Next and three baselines. The first seven benchmarks report accuracy (\%): MMAU and MMSU assess broad understanding, while CochlScene, Nonspeech7k, MELD, NSynth, and DiDiSpeech evaluate acoustic scenes, non-speech vocal sounds, emotion, musical instruments, and speaker gender, respectively. Clotho-Moment and DESED report temporal mean intersection over union (mIoU, \%) for single-interval localization and multi-interval retrieval, respectively. Baseline results are obtained through our API-based evaluation. Higher is better, and the highest value in each row is boldfaced.}
    \label{tab:audio-understanding}
    \small
    \setlength{\tabcolsep}{6pt}
    \begingroup
    \renewcommand{\arraystretch}{1.2}
    \arrayrulecolor{black}
    \resizebox{\textwidth}{!}{
    \begin{tabular}{lcccc}
        \toprule
        Benchmark & Gemini-3.1-Pro & Gemini-3.5-Flash & Seed-2.0-Lite & Qwen-Audio-3.1-ASR-Flash-Next \\
        \midrule
        \rowcolor{QwenPurpleLight}
        \multicolumn{5}{c}{\color{QwenPurple}\bfseries Comprehensive Audio Understanding (Accuracy)} \\
        MMAU          & 80.90 & 77.20 & 78.40 & \textbf{81.60} \\
        MMSU          & \textbf{85.30} & 78.32 & 81.36 & 82.24 \\
        \midrule
        \rowcolor{QwenPurpleLight}
        \multicolumn{5}{c}{\color{QwenPurple}\bfseries Specialized Audio Understanding (Accuracy)} \\
        CochlScene    & 63.90 & 55.07 & 49.82 & \textbf{76.30} \\
        Nonspeech7k   & 90.90 & 76.69 & 88.83 & \textbf{93.20} \\
        MELD          & 52.40 & 45.94 & 57.93 & \textbf{58.12} \\
        NSynth        & 39.20 & 16.31 & 29.79 & \textbf{66.90} \\
        DiDiSpeech    & 93.00 & 64.00 & 98.33 & \textbf{99.67} \\
        \midrule
        \rowcolor{QwenPurpleLight}
        \multicolumn{5}{c}{\color{QwenPurple}\bfseries Temporal Grounding (mIoU)} \\
        Clotho-Moment & 52.56 & 44.24 & 71.79 & \textbf{82.20} \\
        DESED         & 49.37 & 37.86 & \textbf{66.64} & 66.60 \\
        \bottomrule
    \end{tabular}}
    \endgroup
\end{table*}
Qwen-Audio-3.1-ASR-Flash-Next achieves the best performance on seven of the nine benchmarks. On the comprehensive audio understanding benchmarks, it obtains an accuracy of 81.60\% on MMAU, which evaluates audio understanding and reasoning across speech, non-speech sounds, and music, outperforming Gemini-3.1-Pro by 0.70 percentage points. On MMSU, which assesses fine-grained spoken-language perception and reasoning, it achieves an accuracy of 82.24\%, trailing the best result by 3.06 percentage points.

On the specialized audio understanding benchmarks, Qwen-Audio-3.1-ASR-Flash-Next achieves the best results across all five tasks: acoustic scene recognition (CochlScene), non-speech vocal sound recognition (Nonspeech7k), emotion recognition (MELD), musical instrument recognition (NSynth), and speaker gender classification (DiDiSpeech). The improvements are particularly pronounced on CochlScene and NSynth, with absolute margins of 12.40 and 27.70 percentage points over the respective next-best results.

For temporal grounding, Qwen-Audio-3.1-ASR-Flash-Next achieves 82.20\% mIoU on Clotho-Moment, a benchmark for precise single-interval boundary localization, surpassing the next-best result by 10.41 percentage points. On DESED, which evaluates multi-interval retrieval of recurring acoustic events, it achieves 66.60\% mIoU, only 0.04 percentage points below the best result. Together, these results demonstrate robust fine-grained temporal grounding across complementary localization and retrieval settings.

\section{Conclusion}
In this report, we present Qwen-Audio-3.1-ASR, a family of instruction-controlled audio-language models that combines improved speech recognition with speaker-diarized transcription and general audio understanding. Building on a Qwen Mixture-of-Experts backbone and large-scale speech pretraining, the family supports recognition across 30 languages and 16 Chinese dialectal varieties, with user-selectable dialect transcription or translation into Mandarin. It retains production-oriented capabilities including industry-domain entity recognition, historical-context conditioning, hierarchical hotword customization, and native single-pass transcription polishing. The non-streaming Flash and Next variants extend transcription to structured outputs containing speaker labels, timestamps, and recognized text, including overlapping speech. Next additionally supports captioning, temporal grounding, sound event detection, and audio question answering across speech, environmental sounds, and music. Evaluations demonstrate competitive recognition performance across Chinese, English, multilingual, and dialectal benchmarks, together with strong results on speaker-diarized transcription and audio understanding tasks. Deployment variants accommodate asynchronous file transcription, low-latency streaming, and context-aware voice input, with capabilities available according to the selected variant. In particular, Message ASR combines streaming transcript display with an utterance-end full-context refresh, while integrating contextual conditioning, hierarchical hotwords, and native polishing without a separate text-rewriting service. Together, these advances extend the scope of ASR from lexical recognition to speaker-aware transcription and audio-grounded understanding while retaining configurable workflows for practical deployment.

\section{Authors (in Alphabetical Order of Last Name)}

\begin{multicols}{3}
    \begin{itemize}[noitemsep]
        \item Chuanmeng Bian
        \item Daren Chen
        \item Peixin Chen
        \item Zhigao Chen
        \item Chong Deng
        \item Zhiyun Fan
        \item Zhifu Gao
        \item Bo Gong
        \item Qing Gu
        \item Jiajun He
        \item Yawei Hu
        \item Yunjie Ji
        \item Wei Ju
        \item Jingbei Li
        \item Xiangang Li
        \item Xu Li
        \item Zengxi Li
        \item Zheng Li
        \item Chengdong Liang
        \item Baiji Liu
        \item Yaming Liu
        \item Ying Liu
        \item Bin Ma
        \item Yiping Peng
        \item Yuezhang Peng
        \item Zhendong Peng
        \item Yu Pu
        \item Yang Shi
        \item Xin Shu
        \item Jian Tang
        \item Biao Tian
        \item Peiyao Wang
        \item Tianzi Wang
        \item Wen Wang
        \item Wupeng Wang
        \item Cheng Wen
        \item Yuzhong Wu
        \item Zijian Xia
        \item Yunchong Xiao
        \item Nan Yang
        \item Jianwei Yu
        \item Jixing Yu
        \item Binbin Zhang
        \item Lei Zhang
        \item Sitong Zhao
        \item Guangdong Zhou
        \item Yuan Zhou
        \item Jianheng Zhuo
    \end{itemize}
\end{multicols}

\section{Acknowledgments}

We are immensely grateful for the invaluable discussions, support, and assistance provided by many colleagues throughout the development of Qwen-Audio-3.1-ASR. Special thanks go to Qian Chen, Han Zhao, Chaohong Tan, Cheng Wen, Qinglin Zhang, Xiang Lv, Wei Ju, Luyao Cheng, Haoxu Wang, Yongchao He, Chen Ding, Bangduo Chen, Siyu Wang, Zelan Yang, Changwei Li, and Shengkui Zhao.

\bibliographystyle{unsrtnat}
\bibliography{refs}

\end{document}